\documentclass[%
    reprint,
    amsmath,
    amssymb,
    aps,
    pra,
    longbibliography,
    showkeys,
    superscriptaddress,
]{revtex4-2}

\usepackage{graphicx}% Include figure files
\usepackage{dcolumn}% Align table columns on decimal point
\usepackage{hyperref} % Hypertext capabilities

\usepackage{bm}
\usepackage{xcolor} % Text colors
\usepackage[per-mode=symbol]{siunitx} % SI units

\DeclareSIUnit\angstrom{\text{\AA}}         % including Å as a unit
\DeclareSIUnit\elementarycharge{\text{e}}   % including e as a unit
\DeclareSIUnit\atm{\text{atm}}              % including atm as a unit
\DeclareSIUnit\amu{\text{amu}}              % including amu as a unit

\hypersetup{colorlinks=true, 
    linkcolor={blue},
    citecolor={blue}, 
    urlcolor={blue}
}

\begin{document}

\title{Nonlinear phononics through molecular dynamics simulations}

\author{Mike Pols}
\email{mike.pols@mat.ethz.ch}
\affiliation{%
    Materials Theory, ETH Zurich, CH-8093 Z\"urich, Switzerland
}%

\author{Nicola A. Spaldin}
\email{nicola.spaldin@mat.ethz.ch}
\affiliation{%
    Materials Theory, ETH Zurich, CH-8093 Z\"urich, Switzerland
}%

\date{\today}

\begin{abstract}

We present a general method for calculating the dynamics of insulating materials that are driven by THz laser pulses. Our method is based on molecular dynamics (MD) simulations in which, in addition to the usual interatomic interactions, the atoms experience and effective force representing their interaction with the laser pulse. As examples, we demonstrate strong infrared (IR) driving in LiNbO$_{3}$, Raman driving in diamond, and circularly polarized driving in SrTiO$_{3}$. We show that the resulting material dynamics capture the nonlinear responses of these systems under strong THz excitation, as well as the influence that temperature has on these processes. Our findings demonstrate that the presented MD-based method is a viable and convenient way to study nonlinear phononics in materials.

\end{abstract}


\maketitle

\section*{Introduction}

The interaction of strong laser pulses with materials is a powerful method to characterize and control their functionality and structure on ultrafast timescales~\cite{kampfrathResonantNonresonantControl2013}. In recent years, the strong, resonant driving of atomic vibrations in crystals has attracted particular attention as a route for transiently distorting materials away from their equilibrium structure~\cite{forstNonlinearPhononicsUltrafast2011, mankowskyNonequilibriumControlComplex2016, disaEngineeringCrystalStructures2021}. Strongly driving an infrared (IR) or Raman-active phonon mode can induce large vibrational amplitudes and atomic displacements, which can lead to a breakdown of the harmonic approximation. In this regime, the anharmonic terms become increasingly more important, giving rise to a large variety of nonlinear processes that govern the dynamics of the material~\cite{disaEngineeringCrystalStructures2021}. This approach, known as \textit{nonlinear phononics}, has enabled the transient induction of a wide range of materials phenomena, including superconductivity~\cite{faustiLightInducedSuperconductivityStripeOrdered2011, mankowskyNonlinearLatticeDynamics2014}, ferroelectricity~\cite{novaMetastableFerroelectricityOptically2019}, magnetism~\cite{disaPolarizingAntiferromagnetOptical2020}, chirality~\cite{zengPhotoinducedChiralityNonchiral2025}, and the switching of ferroaxiality~\cite{zengPhotoinducedNonvolatileRewritable2025}.

In nonlinear phononics, coupled-oscillator models are widely used to describe the material dynamics~\cite{zeigerTheoryDisplaciveExcitation1992, merlinGeneratingCoherentTHz1997, forstNonlinearPhononicsUltrafast2011, subediTheoryNonlinearPhononics2014, mankowskyNonequilibriumControlComplex2016}. Such models represent the dynamics as a set of interacting phonon modes, typically comprising the driven mode and the modes coupled to it. The interactions between these modes are defined by the potential energy of the system, which is often parameterized in terms of the mode coordinates. Despite their success in describing a wide range of experimentally-observed laser-driven material dynamics~\cite{mankowskyNonlinearLatticeDynamics2014, disaPolarizingAntiferromagnetOptical2020, zengPhotoinducedChiralityNonchiral2025, zengPhotoinducedNonvolatileRewritable2025}, these models have their limitations. Most notably, they can only capture the interactions between modes explicitly included in the model, and since all coupling constants have to be calculated explicitly they typically neglect any coupling away from the zone center~\cite{caruso2025RoadmapUltrafast2026}. Furthermore, finite-temperature effects are generally only included phenomenologically, through effective damping constants.

To overcome these challenges, some studies have started to use out-of-equilibrium molecular dynamics (MD) simulations to simulate the dynamics of materials following THz driving~\cite{shinLightInducedIdealWeyl2024, stoccoElectricfieldDrivenNuclear2025}. In such simulations, the material dynamics are obtained by integrating the equations of motion for the atoms, which experience, in addition to the interatomic forces, an effective force representing the interaction with the laser pulse. These calculations have been enabled by the rapid development of machine-learning interatomic potentials (MLIPs), which can accurately capture higher-order anharmonicity at relatively low computational cost~\cite{verdiThermalTransportPhase2021, ouyangAccurateDescriptionHighorder2022}. In addition, by propagating the quantum dynamics of such driven systems, for instance using the time-dependent self-consistent harmonic approximation (TD-SCHA), quantum nuclear effects can also be taken into account~\cite{libbiNonequilibriumQuantumDynamics2025, libbiAtomisticSimulationsOutofequilibrium2025, libbiLaserdrivenFerroelectricitySrTiO32026}. Despite these advances, the exploration of THz-driven material dynamics has thus far remained mostly limited to linear excitation of IR-active modes, with Raman-based processes and other types of driving, such as circularly polarized laser pulses, remaining largely unexplored.

Here, we introduce a general method for simulating laser-driven dynamics of insulating materials using MD simulations. We describe the potential energy surface of the materials using MLIPs trained on first-principles data, which accurately capture their anharmonic behavior. The electric field of the laser pulse, modeled as a Gaussian pulse, couples to the atoms through the Born effective charges and atomic Raman tensors, which are also calculated through first-principles calculations. By tuning the laser pulse parameters, such as frequency and polarization, we can selectively drive IR- and Raman-active modes. As examples, we show strong nonlinear IR driving in lithium niobate (LiNbO$_{3}$), Raman-based driving in diamond, and circular IR driving in strontium titanate (SrTiO$_{3}$). We release our implementation as the open-source code \texttt{LuMD}~\cite{lumdCode2026}, making the method directly applicable to a broad range of insulating materials. Together, these results introduce a new versatile framework for simulating laser-driven dynamics in materials at finite temperatures.

\section{Theory}

We begin by describing the theory behind the method used to simulate the dynamics of laser-driven materials. In particular, we discuss the laser driving in molecular dynamics simulations and the commonly used phenomenological approach based on coupled-oscillator models.

\subsection{Laser driving}

In molecular dynamics (MD) simulations, the time evolution of a system of interacting atoms (or ions), is determined through Newton's equations of motion, as
\begin{equation}
    \label{eq:newton_eom}
    m_{\kappa} \Ddot{\mathbf{r}}_{\kappa} = \mathbf{F}_{\kappa},
\end{equation}
where $m_{\kappa}$ is the mass of atom $\kappa$, $\Ddot{\mathbf{r}}_{\kappa}$ is its acceleration or the second time derivative of its position, and $\mathbf{F}_{\kappa}$ is the total force acting on it. This force can be decomposed into an internal and external contribution:
\begin{equation}
    \label{eq:atomic_force}
    \mathbf{F}_{\kappa} = \mathbf{F}_{\kappa}^{\mathrm{int}} + \mathbf{F}_{\kappa}^{\mathrm{ext}} = - \nabla_{\kappa} U \left( \{ \mathbf{r}_{\kappa} \} \right) + \mathbf{F}_{\kappa}^{\mathrm{ext}}.
\end{equation}
Here, the internal force is the derivative of the potential energy of the system $U$, a function of the positions of all $N$ particles $\{ \mathbf{r}_{\kappa} \} = \{ \mathbf{r}_{1}, \ldots, \mathbf{r}_{N} \}$, with respect to the position of atom $\kappa$ ($\nabla_{\kappa}$). The external contributions to the atomic force can stem from different origins, such as shear forces and electric or magnetic fields, and can therefore have a variety of forms.

For a laser pulse, the response of a material is primarily dictated by its interaction with the electric field $\mathbf{E}$, which can result in two types of interactions:
\begin{equation}
    \begin{split}
        \label{eq:external_forces}
        \mathbf{F}_{\kappa}^{\mathrm{ext}} = - \frac{\partial U}{\partial \mathbf{r}_{\kappa}} &= \mathbf{F}_{\kappa}^{\mathrm{IR}} + \mathbf{F}_{\kappa}^{\mathrm{Raman}} \\
        &= \mathbf{Z}_{\kappa}^{*} \mathbf{E} + \frac{1}{2} \epsilon_{0} \mathbf{E}^{T} \left( \frac{\partial \bm{\alpha}}{\partial \mathbf{r}_{\kappa}} \right) \mathbf{E}.
    \end{split}
\end{equation}
The first contribution is the infrared (IR) force, $\mathbf{F}_{\kappa}^{\mathrm{IR}}$, which originates from the coupling between the electric field and the dipole moment associated with atom $\kappa$. This term is determined by the Born effective charge tensor $\mathbf{Z}^{*}_{\kappa}$~\cite{gonzeDynamicalMatricesBorn1997, ghosezDynamicalAtomicCharges1998}. The second contribution is the Raman force, $\mathbf{F}_{\kappa}^{\mathrm{Raman}}$, which arises from the field-induced modification of the material polarizability. Here, $\epsilon_{0}$ denotes the vacuum permittivity, $\mathbf{E}^{T}$ is the transpose of the electric field, and $\bm{\alpha}$ is the polarizability tensor. The derivative $\frac{\partial \bm{\alpha}}{\partial \mathbf{r}_{\kappa}}$ describes the variation of the polarizability with respect to the displacement of atom $\kappa$, and it has been referred to as the atomic Raman tensor in the literature~\cite{olearyRapidCharacterizationPoint2024}. A detailed derivation of the forms of these two atomic forces can be found in Note 1 in the Supplemental Material (SM)~\cite{supplementalMaterial} (see also Refs.~\cite{bartokRepresentingChemicalEnvironments2013, parrinelloCrystalStructurePair1980, parrinelloPolymorphicTransitionsSingle1981, togoFirstprinciplesPhononCalculations2023, togoImplementationStrategiesPhonopy2023, gonzeInteratomicForceConstants1994, sirenkoSoftmodeHardeningSrTiO32000} therein).

\subsection{Laser pulse}

The electric field of a laser pulse can be modeled as an oscillating electric field within a Gaussian envelope. As an example, a laser pulse propagating along the $x$ axis with its polarization oriented along the $z$ axis, can be described as
\begin{equation}
    \label{eq:laser_pulse}
    \mathbf{E} ( t ) = \tilde{E}_{0} \exp \left \{ - \frac{(t - t_{0})}{2 \left( \tau / \sqrt{8 \ln 2} \right)} \right \} \begin{bmatrix}
        0 \\
        0 \\
        \cos(\omega_0 t)
    \end{bmatrix},
\end{equation}
where $\tilde{E}_{0}$ is the shielded peak electric field amplitude, $\tau$ the full width at half maximum (FWHM) of the pulse envelope, and $\omega_{0}$ the frequency. The peak electric field within a material is shielded through dielectric screening, reducing the incoming electric field to $\tilde{E}_0 = \frac{2}{1 + \sqrt{\epsilon^{\infty}}} \, E_{0}$, with $E_{0}$ the external electric field and $\epsilon_{\infty}$ the macroscopic dielectric tensor of the material. Different laser pulse polarizations and circular or elliptical pulses, can be modeled by changing the orientation and or phase of the oscillation.

\subsection{Mode dynamics}

The dynamics of the phonon modes in a material can be extracted from the atomic trajectories generated by the MD simulations. By projecting the time-dependent atomic displacements and velocities onto the phonon eigenvectors,
\begin{align}
    \label{eq:normal_mode_mapping_gamma}
    Q_{\mathbf{q} \nu} ( t ) &= \sum_{\kappa} \sqrt{m_{\kappa}} \mathbf{e}^{*}_{\mathbf{q} \nu, \kappa} \cdot \mathbf{u}_{\kappa} ( t ) \, e^{-i \mathbf{q} \cdot \mathbf{r}_{\kappa}^{0}} \\
    \dot{Q}_{\mathbf{q} \nu} ( t ) &= \sum_{\kappa} \sqrt{m_{\kappa}} \mathbf{e}^{*}_{\mathbf{q} \nu, \kappa} \cdot \mathbf{v}_{\kappa} ( t ) \, e^{-i \mathbf{q} \cdot \mathbf{r}_{\kappa}^{0}},
\end{align}
the lattice dynamics can be decomposed into contributions from individual phonon modes. Here, $\mathbf{e}_{\mathbf{q} \nu, \kappa}$ denotes the phonon eigenvector of mode $\nu$ at wave vector $\mathbf{q}$, and $\mathbf{r}_{\kappa}^{0}$ is the equilibrium position of atom $\kappa$. The time-dependent displacement, $\mathbf{u}_{\kappa} ( t ) = \mathbf{r}_{\kappa} ( t ) - \mathbf{r}_{\kappa}^{0}$, and velocity, $\mathbf{v}_{\kappa} ( t )$, are obtained from the MD trajectory. This mass-weighted projection transforms the atomic motion from real space into the phonon normal-mode representation, yielding the time-dependent mode amplitude $Q_{\mathbf{q} \nu} ( t )$ and its corresponding velocity $\Dot{Q}_{\mathbf{q} \nu} ( t )$.

\subsection{Phenomenological models}

Since we will compare the dynamics from laser-driven MD simulations with those described by phenomenological models, we briefly review how the latter can be used to describe material dynamics. Such models represent the time evolution of the relevant modes through coupled oscillators, and, given the potential energy surface $U$ as a function of the mode amplitudes, their dynamics are governed by the equations of motion
\begin{equation}
    \label{eq:equations_of_motion}
    \Ddot{Q}_{\nu} + \gamma_{\nu} \Dot{Q}_{\nu} + \frac{\partial U}{\partial Q_{\nu}} = \mathbf{Z}_{\nu} \cdot \mathbf{E} + \frac{1}{2} \epsilon_{0} \mathbf{E}^{T} \mathbf{R}_{\nu} \mathbf{E}.
\end{equation}
Here, $\Ddot{Q}_{\nu}$, $\Dot{Q}_{\nu}$, $Q_{\nu}$, and $\gamma_{\nu}$ are the acceleration, velocity, amplitude, and damping coefficient (or linewidth) of mode $\nu$, respectively. The anharmonic terms in $U$ introduce couplings between different normal modes, enabling energy transfer between them.

The right-hand side of the equation of motion describes the driving force exerted on a mode by an external electric field, with the time dependence of the field given by Eq.~\ref{eq:laser_pulse}. The first term represents the coupling between the electric field and the mode effective charge, leading to IR-based excitation of the mode. The mode effective charge is defined as
\begin{equation}
    \mathbf{Z}_{\nu} = \sum_{\kappa} \mathbf{Z}^{*}_{\kappa} \frac{\mathbf{e}_{\mathbf{q} \nu, \kappa}}{\sqrt{m_{\kappa}}}.
\end{equation}

The second term describes the coupling between the electric field and the mode Raman tensor, corresponding to excitation through the electric-field-induced modification of the polarizability. The Raman tensor is given by
\begin{equation}
    \mathbf{R}_{\nu} = \sum_{\kappa} \left( \frac{\partial \bm{\alpha}}{\partial \mathbf{r}_{\kappa}} \right) \frac{\mathbf{e}_{\mathbf{q} \nu, \kappa}}{\sqrt{m_{\kappa}}}.
\end{equation}
By numerically integrating these equations of motion, the phenomenological model captures the time-dependent evolution of coherently driven modes and the energy transfer between coupled modes mediated by anharmonic interactions.

\section{Computational details}

Next, we detail the computational settings used in our laser driving simulations. Among others, we specify the parameters used for the first-principles calculations for the material properties, the training of the machine-learning interatomic potentials, and the molecular dynamics simulations.

\subsection{First-principles calculations}

We calculate the material properties with density functional theory (DFT) as implemented in the Vienna Ab-initio Simulation Package (\texttt{VASP})~\cite{kresseInitioMolecularDynamics1993, kresseInitioMoleculardynamicsSimulation1994, kresseEfficiencyAbinitioTotal1996, kresseEfficientIterativeSchemes1996}. Projector-augmented wave (PAW) pseudopotentials~\cite{kresseUltrasoftPseudopotentialsProjector1999} are used, with the following valence electron configurations (name of pseudopotentials): 2s$^{2}$ (\texttt{Li}), 2s$^{2}$2p$^{2}$ (\texttt{C}), 2s$^{2}$2p$^{4}$ (\texttt{O}), 3s$^{2}$3p$^{6}$3d$^{3}$4s$^{1}$ (\texttt{Ti\_sv}), 4s$^{2}$4p$^{6}$5s$^{2}$ (\texttt{Sr}), and 4p$^{6}$5s$^{1}$4d$^{4}$ (\texttt{Nb\_pv}). The exchange-correlation interaction is modeled within the generalized gradient approximation using the PBEsol functional~\cite{perdewRestoringDensityGradientExpansion2008}. For LiNbO$_{3}$ and SrTiO$_{3}$ a cutoff energy of \SI{600}{\eV} is used, while for diamond it is set to \SI{950}{\eV}. The reciprocal space is sampled with a $\mathbf{k}$-mesh of $6 \times 6 \times 6$ for LiNbO$_{3}$, $12 \times 12 \times 12$ for SrTiO$_{3}$, and $9 \times 9 \times 9$ for diamond. The crystal structures are optimized by relaxing the atomic positions, cell shape, and cell volume until the total energy converges to within $10^{-6}~\SI{}{\meV}$ and the atomic forces converge to within $10^{-2}~\SI{}{\meV\per\angstrom}$.

The macroscopic dielectric tensor and Born effective charges are then calculated for the optimized structures using density functional perturbation theory (DFPT). The atomic Raman tensors are extracted from the dielectric tensors using a finite differences method, with the relation: $\frac{\partial \alpha}{\partial \mathbf{r}_{\kappa}} = \epsilon_{0} V \frac{\partial \epsilon^{\infty}}{\partial \mathbf{r}_{\kappa}}$, where $\epsilon_{0}$ is the vacuum permittivity and $V$ the unit cell volume. We find that atomic displacements of \SI{0.15}{\angstrom} from the equilibrium position are sufficient to converge the atomic Raman tensors. Additional details on the calculation of the atomic Raman tensors and its convergence with respect to the atomic displacements can be found in the SM Note 2~\cite{supplementalMaterial}.

\subsection{Machine-learning interatomic potentials}

To investigate the laser-driven material dynamics, we train machine-learning interatomic potentials (MLIPs) on DFT data. The reference data are generated through on-the-fly molecular dynamics (MD) simulations, using the structural similarity kernel as implemented in \texttt{VASP}. For each material, we perform multiple $NpT$ simulations of supercells at elevated temperatures and under atmospheric pressure, ensuring a sufficient sampling of the structural distortions and accompanying anharmonicity of the lattice. For LiNbO$_{3}$ and SrTiO$_{3}$ we run simulations at \SI{500}{\K}, \SI{1000}{\K}, and \SI{1500}{\K}, while we sample the reference data for diamond at \SI{500}{\K}, \SI{1500}{\K}, and \SI{2500}{\K}, to capture the behavior of its high-frequency modes.

The generated datasets, containing energies, forces, and stresses, are then used to train separate MACE models for each material~\cite{batatiaMACEHigherOrder2022, batatiaDesignSpaceE3equivariant2025}. These \texttt{ScaleShiftMACE} models have two message passing layers, 64 equivariant messages, and a correlation order of 3. For LiNbO$_{3}$ and SrTiO$_{3}$ we use a cutoff radius of \SI{5.0}{\angstrom}, which we decrease to \SI{3.0}{\angstrom} for diamond due to its higher atomic density. We train for 1000 epochs with a batch size of 5. As the loss function we use a Huber loss function with weights 1, 100, and 0.1, for the energies, forces, and stresses, respectively. For the final 25\% of epochs, the weights are increased to 1000, 100, and 10. Details on the dataset generation, training procedure, and model validations can be found in the SM Note 3~\cite{supplementalMaterial}.

\subsection{Molecular dynamics}

The trained MLIPs are used for laser-driven MD simulations in \texttt{LAMMPS}~\cite{thompsonLAMMPSFlexibleSimulation2022}. Before the production runs, the system geometry is optimized and equilibrated to the target temperature and atmospheric pressure through successive $NpT$ and $NVT$ simulations, each lasting \SI{10.0}{\ps}. The time-averaged equilibrium structure is then sampled during an additional \SI{5.0}{\ps} $NVT$ simulation. Subsequently, the laser pulse is applied in an $NVT$ simulation of \SI{15.0}{\ps}, with the pulse centered at \SI{5.0}{\ps}. To account for the stochastic nature of MD, the laser-driven simulations are repeated with different initial conditions, yielding an ensemble of trajectories. The material response is obtained by averaging over this ensemble.

During the simulations, the temperature and pressure are controlled using Nos\'{e}-Hoover-style thermostat and barostat algorithms, respectively~\cite{shinodaRapidEstimationElastic2004, parrinelloPolymorphicTransitionsSingle1981, tuckermanLiouvilleoperatorDerivedMeasurepreserving2006}. In all simulations, we use a chain length of 3 with a damping constant of $\tau_{T} = \SI{100}{\fs}$ for the thermostat and $\tau_{p} = \SI{1000}{\fs}$ for the barostat. The exception is the laser-driving step, where the thermostat damping constant is increased to $\tau_{T} = \SI{2000}{\fs}$. In all simulations a time step of \SI{1.0}{\fs} is used for the time integration~\cite{tuckermanLiouvilleoperatorDerivedMeasurepreserving2006}.

\section{Applications}

\begin{figure}[htbp!]
    \includegraphics{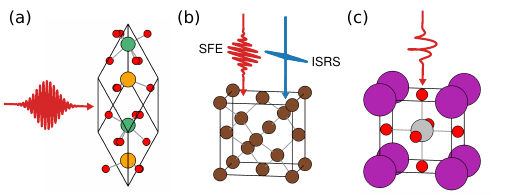}
    \caption{Examples of laser driving studied in this work. (a) Infrared (IR) driving of LiNbO$_{3}$. (b) Raman driving of diamond through sum-frequency excitation (SFE) and impulsive stimulated Raman scattering (ISRS). (c) Circular IR driving of SrTiO$_{3}$. Atoms have the following colors: Li (orange), Nb (green), O (red), C (brown), Sr (purple), and Ti (gray). Structures are visualized using VESTA~\cite{mommaVESTA3Threedimensional2011}.}
    \label{fig:materials_overview}
\end{figure}

In the following, we demonstrate how our method can be used to study the laser-driven phonon dynamics of representative insulating materials at finite temperatures. Using LiNbO$_{3}$ [Fig.~\ref{fig:materials_overview}(a)], diamond [Fig.~\ref{fig:materials_overview}(b)], and SrTiO$_{3}$ [Fig.~\ref{fig:materials_overview}(c)], we illustrate infrared (IR) and Raman driving in materials, with both linearly and circularly polarized laser pulses (SM Note 4~\cite{supplementalMaterial}).

\subsection{Lithium niobate}

LiNbO$_{3}$ is a ferroelectric material with crystal structure in the $R3c$ space group~\cite{abrahamsFerroelectricLithiumNiobate1966}. In this symmetry, the phonon modes have $A_{1}$, $A_{2}$, or $E$ irreducible representations, of which the $A_{1}$ and $E$ modes are both IR- and Raman-active, making it a popular subject for experimental studies of lattice dynamics and nonlinear phononics~\cite{mankowskyUltrafastReversalFerroelectric2017, vonhoegenProbingInteratomicPotential2018, henstridgeNonlocalNonlinearPhononics2022, biggsUltrafastFaradayRotation2026, biggsMeasuringMomentumresolvedDissipation2026}. In particular, the highest-frequency $A_{1}$ mode at approximately \SI{19}{\THz} has attracted considerable attention as a means of inducing ferroelectric switching through its anharmonic coupling to the low-frequency ferroelectric $A_{1}$ mode at approximately \SI{7}{\THz}~\cite{subediProposalUltrafastSwitching2015, mankowskyUltrafastReversalFerroelectric2017, abalmasovUltrafastReversalFerroelectric2020, henstridgeNonlocalNonlinearPhononics2022}.

\begin{figure}[htbp]
    \includegraphics{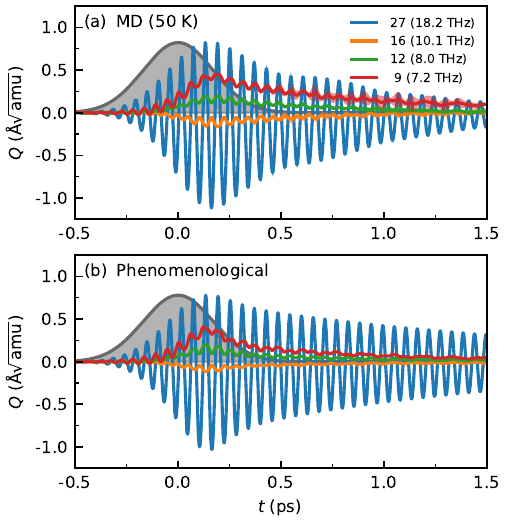}
    \caption{Calculated dynamics of laser-driven LiNbO$_{3}$. Mode amplitudes of $A_{1}$ modes driven with an \SI{18.2}{\THz} laser pulse for (a) molcular dynamics (MD) simulation for a system of 2880 atoms at \SI{50}{\K} and (b) a phenomenological model at \SI{0}{\K}. The dynamics from MD are averaged dynamics over five simulations, with the 95\% (2$\sigma$) confidence interval for the mode amplitudes indicated in shaded areas. The envelope of the laser pulse is indicated in gray.}
    \label{fig:linbo3_md_phenomenological}
\end{figure}

To resonantly excite this high-frequency $A_{1}$ mode in our simulations, we use a laser pulse that is linearly polarized along the ferroelectric $z$ axis, with a frequency of \SI{18.2}{\THz}. The pulse width and peak electric field are set to $\tau = \SI{0.4}{\ps}$ and $E_{0} = \SI{8.0}{\mega\volt\per\cm}$. The resulting dynamics of the $A_{1}$ modes are shown in Fig.~\ref{fig:linbo3_md_phenomenological}(a). As the laser pulse arrives, the high-frequency mode 27, with a frequency of \SI{18.2}{\THz}, is excited. Interestingly, we also observe a finite amplitude developing in the other three $A_{1}$ modes, i.e. mode 9, 12, and 16. We emphasize that these additional modes are not driven by the laser pulse, but instead attain a finite amplitude through nonlinear mode coupling. Specifically, the relevant anharmonic interaction for this is the $U^{\mathrm{anh}} \sim Q_{27}^{2} Q_{a}$ term~\cite{subediProposalUltrafastSwitching2015, mankowskyUltrafastReversalFerroelectric2017}, where $Q_{27}$ denotes the mode amplitude of the high-frequency mode and $Q_{a}$ that of the lower frequency $A_{1}$ modes. As a result of this coupling, the strongly driven mode at \SI{18.2}{\THz} exerts a net force on the other $A_{1}$ modes, rectifying their amplitudes.

When we compare the mode dynamics from our MD simulations to those of the phenomenological model [Fig.~\ref{fig:linbo3_md_phenomenological}(b)], we see that both methods yield rather similar overall dynamics, with the main feature being a rectification of the other $A_{1}$ modes by the driven mode. The similarity between the results of the two methods demonstrates that the MLIP accurately reproduces the mode couplings, which we also confirm in SM Note 3~\cite{supplementalMaterial}. Beyond this broad agreement, we see that the MD-calculated dynamics exhibit additional features absent from the phenomenological model. First, the MD simulations display thermal fluctuations in the mode amplitudes, which stem from the finite-temperature nature of the simulations. Moreover, the driven mode exhibits a clear non-smooth decay of its amplitudes, indicating that its dynamics are governed by complex energy-transfer processes involving a large number of modes. We emphasize that these physical features are not easily captured by a phenomenological model, motivating the use of MD-based calculations.

\begin{figure}[htbp!]
    \includegraphics{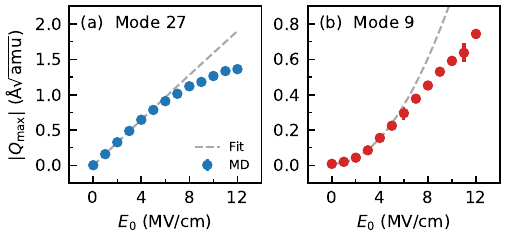}
    \caption{MD-calculated field scaling of laser-driven mode dynamics in LiNbO$_{3}$. Maximum mode amplitudes $| Q^{\mathrm{max}} |$ of (a) mode 27 and (b) mode 9 as a function of the peak electric field for an \SI{18.2}{\THz} laser pulse in molecular dynamics (MD) simulations. The amplitudes are averaged over five independent MD simulations, with the error bars indicating the 95\% ($2 \sigma$) confidence interval; in many cases the error bars are smaller than the marker size. The dashed gray line indicates a linear and quadratic fit for mode 27 and mode 9, respectively.}
    \label{fig:linbo3_field_scaling}
\end{figure}

Next, to demonstrate that mode 27 is excited through IR-driving and not Raman-driving, we investigate how its amplitude scales with the peak electric field $E_{0}$. Keeping all other pulse parameters fixed, we vary the peak electric field from 0.0 to \SI{12.0}{\mega\volt\per\cm}. The resulting maximum amplitudes $| Q^{\mathrm{max}} |$ of mode 27 and mode 9 are shown in Fig.~\ref{fig:linbo3_field_scaling}. As shown in Fig.~\ref{fig:linbo3_field_scaling}(a), at low field strengths, below \SI{4.0}{\mega\volt\per\cm}, the maximum amplitude of the driven mode increases linearly with $E_{0}$. This behavior is expected for an IR-driven mode, for which the electric field generates an effective force $F_{27}^{\mathrm{IR}} \sim \mathbf{Z}_{27} \cdot \mathbf{E}$, that is linear in the electric field. In contrast, the maximum amplitude of mode 9, shown in Fig.~\ref{fig:linbo3_field_scaling}(b), exhibits a quadratic dependence on the peak electric field. This behavior can be understood from the effective driving force exerted on mode 9 through its coupling to mode 27, given by $F_{9}^{\mathrm{coupling}} = -\frac{\partial U^{\mathrm{anh}}}{\partial Q_{9}} \sim - Q_{27}^{2}$. Since the amplitude of mode 27 scales linearly with $E_{0}$ in this regime, the coupling force acting on mode 9 consequently scales as $E_{0}^{2}$, resulting in the observed quadratic dependence.
At larger peak electric fields, the maximum amplitude of mode 27 deviates from the linear dependence observed at low field strengths. Correspondingly, the quadratic dependence of mode 9 also breaks down at larger electric fields. Both deviations indicate that for these pulse strengths the response of LiNbO$_{3}$ is no longer well-described by the linear-response regime, and that higher-order anharmonic effects become increasingly more important as the amplitude of the driven mode grows.

\begin{figure}[htbp]
    \includegraphics{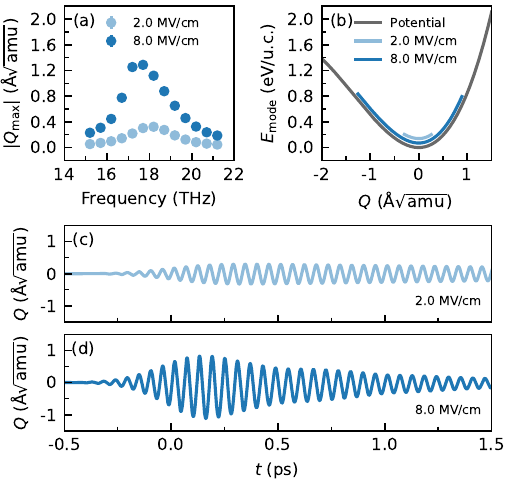}
    \caption{Resonant driving of mode 27 of LiNbO$_{3}$. (a) Resonance curves for laser pulses with different peak electric fields. (b) Mode dynamics within mode potential energy. Mode amplitudes resulting from an \SI{18.2}{\THz} laser pulse with a peak electric field of (c) \SI{2.0}{\mega\volt\per\cm} and (d) \SI{8.0}{\mega\volt\per\cm}.}
    \label{fig:linbo3_resonance}
\end{figure}

We now turn to the frequency dependence of the mode dynamics to better understand the origin of this nonlinear behavior. By varying the central frequency of the laser pulse, we can directly probe the proximity of the driven mode to resonance, taking  into account the dependence of the resonance frequency on the driving intensity. Specifically, we vary the frequency of the laser pulse from \SI{15.2}{\THz} to \SI{21.2}{\THz} in steps of \SI{0.5}{\THz} and consider peak electric fields of \SI{2.0}{\mega\volt\per\cm} and \SI{8.0}{\mega\volt\per\cm}. The resonance curves and mode dynamics for mode 27, that result from this analysis, are shown in Fig.~\ref{fig:linbo3_resonance}.

For a peak electric field of \SI{2.0}{\mega\volt\per\cm}, the resonance curve is symmetric and centered at \SI{18.2}{\THz} [Fig.~\ref{fig:linbo3_resonance}(a)]. This is consistent with the linear-response regime identified above. At this field strength, the amplitude of the drive oscillation remains small, meaning that the mode only samples the region close to the minimum of the potential energy landscape [Fig.~\ref{fig:linbo3_resonance}(b)], where it is well-approximated by a harmonic potential. Therefore, the driven mode oscillates at the harmonic frequency of the mode, i.e. \SI{18.2}{\THz}.

By increasing the peak electric field to \SI{8.0}{\mega\volt\per\cm}, mode 27 is driven to substantially larger amplitudes. As a result, the mode samples a much broader region of the energy landscape [Fig.~\ref{fig:linbo3_resonance}(b)], which now also includes the anharmonic terms, such as third-order contributions, resulting in a breakdown of the harmonic approximation. Consequently, under strong driving, the mode no longer oscillates at its harmonic frequency, but instead at a shifted frequency. The resonance curve in Fig.~\ref{fig:linbo3_resonance}(a) shows that for $E_{0} = \SI{8.0}{\mega\volt\per\cm}$, the driven mode has a frequency of approximately \SI{17.5}{\THz}, which, compared to the harmonic case, is softer by about \SI{0.7}{\THz}.

The above observation can be explained by looking at the time-dependent mode amplitudes for these two peak electric fields in Figs.~\ref{fig:linbo3_resonance}(c),(d). In the linear regime, at \SI{2.0}{\mega\volt\per\cm}, the mode has equal positive and negative displacements, oscillating symmetrically with small perturbations about its equilibrium position [Fig.~\ref{fig:linbo3_resonance}(c)]. In contrast, under strong driving (\SI{8.0}{\mega\volt\per\cm}), the oscillation of the mode is no longer symmetric, with the mode reaching larger negative than positive amplitudes [Fig.~\ref{fig:linbo3_resonance}(d)]. As shown in Fig.~\ref{fig:linbo3_resonance}(d), the mode explores a much larger region of the potential energy surface, particularly its anharmonic parts. Since the leading anharmonic term in the potential energy of the mode is third order in the mode amplitude, we observe the mode frequency shift to lower values.

Recently, the same third-order anharmonic term has also been connected to \textit{ferrons}, which are excitations of the ferroelectric order carrying a net dipole moment~\cite{tangExcitationsFerroelectricOrder2022, bauerPolarizationTransportFerroelectrics2023, tangElectricAnalogMagnons2024}. As a result of the asymmetry under strong driving [Fig.~\ref{fig:linbo3_resonance}(b) and Fig.~\ref{fig:linbo3_resonance}(d)], the driven mode also yields a transiently induced net dipole moment upon excitation. Therefore, similar to other light-induced ferroelectric excitations~\cite{polsLatticeExcitationsFinite2026}, the asymmetric component of the \SI{18.2}{\THz} mode can also be classified as a ferron.

Apart from differences in the amplitude and symmetry of the mode dynamics, the time evolution also shows that weak and strong driving lead to different temporal responses. Whereas we barely observe any decay within the first \SI{1.5}{\ps} for weak driving, the mode amplitude decays substantially when strongly driven. This decay stems from coupling between phonon modes, which, as confirmed earlier [Fig.~\ref{fig:linbo3_field_scaling}], becomes stronger with the driving strength. Furthermore, we highlight that the non-smooth decay that we observed earlier and connected to complex energy transfer between modes, is more apparent and stronger under strong driving.

\begin{figure}[htbp]
    \includegraphics{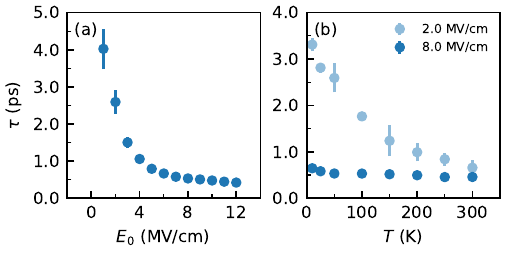}
    \caption{Phonon lifetimes of driven mode in LiNbO$_{3}$. (a) Dependence of phonon lifetime on peak electric field $E_{0}$ of the laser pulse. (b) Temperature dependence of the phonon lifetime. }
    \label{fig:linbo3_lifetimes}
\end{figure}

For a better understanding of the coupling and energy transfer between driven modes, we now analyze how the lifetime of the modes is affected by the driving strength and temperature. The lifetimes, determined from the autocorrelation function of the amplitude of the driven mode (see SM Note 5~\cite{supplementalMaterial} for details), are shown as a function of peak electric field and temperature in Fig.~\ref{fig:linbo3_lifetimes}.

Figure~\ref{fig:linbo3_lifetimes}(a) illustrates that the stronger a mode is driven, the smaller its lifetime becomes. This is consistent with our earlier observation of the mode amplitudes, where an appreciable decay shortly after the laser pulse is only seen in strongly driven systems. We stress the decrease in the lifetime of the phonon mode stems from a stronger anharmonic coupling experienced by modes that are driven to large amplitudes. Notably, this effect is reduces the lifetime by an order of magnitude from \SI{4.0}{\ps} to \SI{0.4}{\ps} for \SI{1.0}{\mega\volt\per\cm} and \SI{12.0}{\mega\volt\per\cm}, respectively.

If, instead of $E_{0}$, the temperature of the simulation is varied, we see a similar effect: the higher the temperature, the shorter the phonon lifetime [Fig.~\ref{fig:linbo3_lifetimes}(b)]. We perform this temperature increase for both the weakly driven (\SI{2.0}{\mega\volt\per\cm}) and strongly driven (\SI{8.0}{\mega\volt\per\cm}) system, with both showing similar trends. At higher temperatures, a larger number of phonon modes become populated, opening up additional decay channels for the driven mode and enhancing the their thermal scattering, which shortens the lifetime of the mode. We note, however, that the effect is much more pronounced in the weakly driven system, where the lifetime drops from \SI{3.3}{\ps} to \SI{0.6}{\ps}, compared to a decrease from \SI{0.6}{\ps} to \SI{0.4}{\ps} in the strongly driven case. This suggests that in the strongly driven regime the decay is already dominated by nonlinear processes and additional thermal scattering has only a small effect on the lifetime of the driven mode.

\subsection{Diamond}

Next we illustrate the driving of material vibrations through Raman-active phonons, using the example of the carbon allotrope diamond, which is commonly studied in the context of phonon excitations through Raman-based processes~\cite{ishiokaCoherentOpticalPhonons2006, maehrleinTerahertzSumFrequencyExcitation2017, juraschekSumfrequencyIonicRaman2018}. Due to its highly symmetric $Fd\bar{3}m$ crystal structure~\cite{braggStructureDiamond1913}, its zone-center optical phonon \SI{39}{\THz} is Raman active and transforms according to the three-dimensional $T_{2g}$ irreducible representation. The three components of this representation correspond to distortions transforming like the quadratic functions $xy$, $zx$, and $yz$, which are degenerate due to the cubic symmetry of the material.

\begin{figure}[htbp]
    \includegraphics{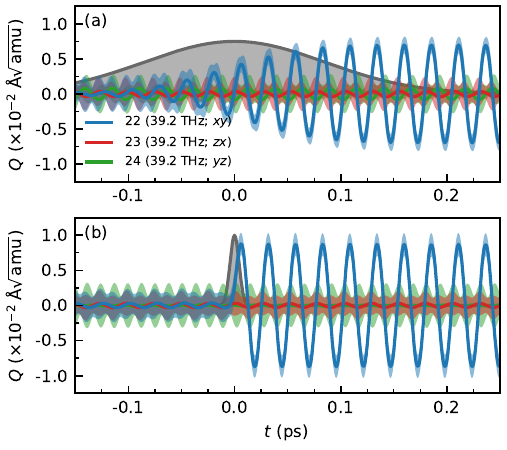}
    \caption{MD-calculated dynamics of laser-driven diamond. Mode amplitude of $T_{\mathrm{2g}}$ mode driven at \SI{50}{\K} with (a) \SI{19.5}{\THz} laser pulse via sum-frequency excitation (SFE) and (b) \SI{759}{\THz} laser pulse via impulsive stimulated Raman scattering (ISRS). Mode 22, 23, 24 correspond to the $xy$, $zx$, and $yz$ components of the $T_{2g}$ mode, respectively. Dynamics are averaged over ten simulations, with the 95\% (2$\sigma$) confidence interval for the mode amplitudes indicated in shaded areas. The envelope of the laser pulse is indicated in gray.}
    \label{fig:diamond_raman_driving}
\end{figure}

Here, we excite the Raman-active mode at \SI{39}{\THz} in diamond via two types of Raman processes. Unlike IR-active modes, Raman-active modes can not be directly driven at the frequency of a laser pulse. Because the driving force of Raman excitations scales with the square of the electric field $E^{2}$, the effective driving force oscillates at twice the frequency. Consequently, a Raman-active mode can only be resonantly excited by a laser pulse with a frequency that is half that of the mode, a process that is commonly referred to as sum-frequency excitation (SFE). Therefore, we use a laser pulse with a frequency of \SI{19.5}{\THz} ($\tau = \SI{0.2}{\ps}$; $E_{0} = \SI{32}{\mega\volt\per\cm}$), polarized along the $xy$ axis, to drive the \SI{39}{\THz} mode via SFE in our simulations. The resulting dynamics of the $T_{2g}$ mode are shown in Fig.~\ref{fig:diamond_raman_driving}(a). As for IR driving, the amplitude of the mode that is excited by SFE grows gradually while the laser pulse irradiates the sample, however, in contrast to the IR case, here the driven mode oscillates at twice the laser frequency. Moreover, we observe that only mode 22, the $xy$ component of the $T_{2g}$ mode, is excited, as a consequence of the polarization of the laser pulse. As shown in Fig. S11 in SM Note 6~\cite{supplementalMaterial}, the other components can be excited by rotating the laser pulse polarization to the $zx$ or $yz$ axis.

Besides resonant driving, Raman-active modes can also be excited impulsively, through a process called impulsive stimulated Raman scattering (ISRS). In ISRS, a mode is not driven by twice the frequency of the laser pulse, but instead by the square of the pulse envelope, in other words the intensity profile. A sufficiently short laser pulse can therefore act as an impulsive kick, with the short duration of the pulse ensuring that the frequency of the mode of interest is covered by the broad spectral bandwidth. To excite the $T_{2g}$ mode with ISRS, we use an $xy$-oriented laser pulse with a frequency of \SI{759}{\THz} and a width of $\tau = \SI{10}{\fs}$ ($E_{0} = \SI{128}{\mega\volt\per\cm}$), of which the spectral bandwidth also covers the \SI{39}{\THz} (see SM Note 4~\cite{supplementalMaterial}). Looking at the dynamics that result from this excitation [Fig.~\ref{fig:diamond_raman_driving}(b)], we observe that the short laser pulse impulsively excites the $T_{2g}$ mode, as demonstrated by the sharp rise of the amplitude within a fraction of the mode period. As was the case for the SFE excitation, only the $xy$ component of the $T_{2g}$ mode is excited, due to the polarization of the short laser pulse.

Comparing the amplitudes of the modes driven by Raman-based processes [Fig.~\ref{fig:linbo3_md_phenomenological}(a)] with those of IR-driven modes [Fig.~\ref{fig:diamond_raman_driving}], we observe that the maximum amplitude reached by Raman-driven modes is about two orders of magnitude smaller than that typically seen for IR-active modes. As a result of this, the amplitudes are much closer to the thermal noise of the modes, which makes Raman-driven mode oscillations considerably harder to resolve than their IR-driven counterparts. This difference stems from the substantially smaller effective driving force experienced by Raman-active modes, which arises from the small $\epsilon_{0}$ prefactor.

\begin{figure}[htbp]
    \includegraphics{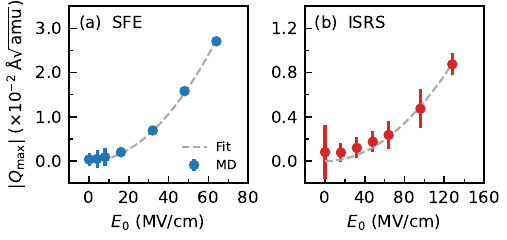}
    \caption{MD-calculated field scaling of laser-driven mode dynamics in diamond. Maximum mode amplitudes $| Q^{\mathrm{max}} |$ of Raman-active modes in diamond as a result of (a) sum-frequency excitation (SFE) and (b) impulsive stimulated Raman scattering (ISRS). The dashed gray line indicates a quadratic fit for both excitation mechanisms.}
    \label{fig:diamond_field_scaling}
\end{figure}

\begin{figure*}[htbp]
    \includegraphics{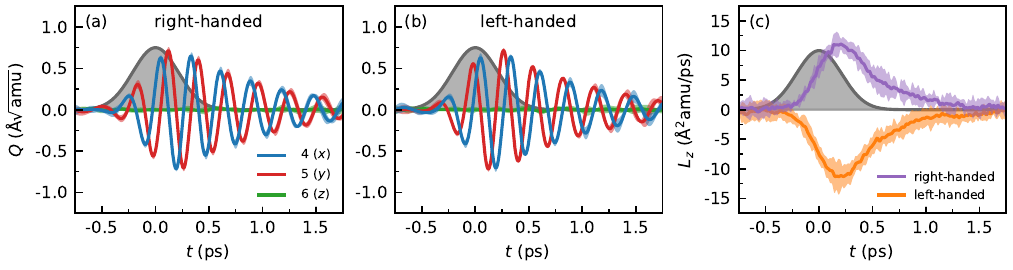}
    \caption{MD-calculated dynamics of circularly laser-driven SrTiO$_{3}$. Mode amplitudes of $T_{1u}$ modes driven at \SI{300}{\K} with (a) right-handed and (b) left-handed circularly polarized laser pulse with a frequency of \SI{3.0}{\THz}. (c) Angular momentum along the $z$ axis in circularly driven SrTiO$_{3}$. Mode 4, 5, and 6 correspond to the $x$, $y$, and $z$ components of the $T_{1u}$ modes, respectively. Dynamics are averaged over five simulations, with the 95\% ($2 \sigma$) confidence interval for the mode amplitudes and angular momentum indicated in the shaded ares. The envelope of the laser pulse is indicated in gray.}
    \label{fig:srtio3_circular_drive}
\end{figure*}

To confirm that the laser excitations shown in Fig.~\ref{fig:diamond_raman_driving} proceed through a Raman process, we finally investigate how the amplitude of the driven mode scales with the peak field of the laser pulse. For both excitation mechanisms, we keep the frequency of the pulse fixed while varying the peak electric field. For SFE, $E_{0}$ is varied up to \SI{64}{\mega\volt\per\cm}, whereas for ISRS it is changed up to \SI{128}{\mega\volt\per\cm}. The maximum amplitude of the driven mode $| Q^{\mathrm{max}} |$ is shown for both mechanisms in Fig.~\ref{fig:diamond_field_scaling}. We see for both mechanisms that the maximum amplitude of the driven mode scales quadratically with the peak field $E_{0}$. This behavior is consistent with Raman excitation of the modes, for which the effective driving force is given by $F^{\mathrm{Raman}}_{22} \sim \frac{1}{2} \epsilon_{0} \mathbf{E}^{T} \mathbf{R}_{22} \mathbf{E}$, which is quadratic in the electric field.

\subsection{Strontium titanate}

Finally, we illustrate laser driving with circularly polarized laser pulses, using the example of SrTiO$_{3}$. At room temperature, SrTiO$_{3}$ has a cubic crystal structure with the $Pm\bar{3}m$ space group~\cite{megawCrystalStructureDouble1946, cowleyLatticeDynamicsPhase1964}. As a result of this symmetry, the zone-center phonon modes can be classified according to the cubic point group. In particular, the lowest frequency optical mode, responsible for the quantum paraelectric behavior at low temperatures~\cite{mullerSrTiO3IntrinsicQuantum1979, zhongEffectQuantumFluctuations1996, rowleyFerroelectricQuantumCriticality2014, shinQuantumParaelectricPhase2021, essweinFerroelectricQuantumParaelectric2022, verdiQuantumParaelectricityStructural2023}, transforms according to the three-dimensional $T_{1u}$ irreducible representation and is also IR-active. The three components of this representation correspond to polar displacements along the three Cartesian directions, i.e. $x$, $y$, and $z$, and are degenerate due to the cubic symmetry. As a result, two in-plane components of this soft mode can be driven with a controlled relative phase, making SrTiO$_{3}$ an ideal system for studying circularly polarized phonons in materials, as has been demonstrated experimentally~\cite{basiniTerahertzElectricfielddrivenDynamical2024, mankowskyQuantifyingAngularMomentum2026, choiLightDrivenUltrafastControl2026, choiTHzDrivenQuantumIonic2026}.

Here, we drive the $T_{1u}$ modes of SrTiO$_{3}$ using laser pulses with the electric field circularly polarized in the $xy$ plane. For the right-handed pulse, the $x$ component of the electric field leads the $y$ component by $\pi/2$, whereas it lags behind by $\pi/2$ for a left-handed pulse. To be close to resonance with the driven modes at \SI{300}{\K} (Fig. S13; SM Note 7~\cite{supplementalMaterial}), we use a frequency of \SI{3.0}{\THz}. The remaining pulse parameters are a peak electric field of $E_{0} = \SI{2.0}{\mega\volt\per\cm}$ and a pulse width of $\tau = \SI{0.5}{\ps}$. The resulting material dynamics at \SI{300}{\K} are shown in Fig.~\ref{fig:srtio3_circular_drive}.

Focusing first on the mode dynamics [Figs.~\ref{fig:srtio3_circular_drive}(a),(b)], we see that the laser pulses indeed only excite the $T_{1u}$ modes that are IR active along the $x$ and $y$ axes, i.e. mode 4 and 5. Analogous to the electric field of the laser pulses, the mode with $x$ activity (mode 4) leads the $y$-active mode (mode 5) by a $\pi/2$ phase shift for the right-handed pulse [Fig.~\ref{fig:srtio3_circular_drive}(a)], whereas it lags behind for the left-handed pulse [Fig.~\ref{fig:srtio3_circular_drive}(b)]. The driven modes grow in amplitude during the laser pulse, reaching their maximum approximately \SI{0.2}{\ps} after the pulse center. Following this, the modes decrease in amplitude as a result of coupling to the other modes, which gradually brings the system back to an equilibrium phonon distribution.

Next, we quantify the angular momentum generated by the laser pulse in the model system. It is of particular interest for circularly polarized phonons, as it characterizes their rotational motion and provides a direct measure for the handedness of the driven dynamics. Since we have access to the full trajectories of the laser-driven dynamics, including the atomic displacements and velocities, we can calculate the total angular momentum $\mathbf{L}$ as
\begin{equation}
    \label{eq:angular_momentum}
    \mathbf{L} ( t ) = \frac{1}{N_{\mathrm{cells}}} \sum_{\kappa} m_{\kappa} \mathbf{u}_{\kappa} ( t ) \times \mathbf{v}_{\kappa} ( t ),
\end{equation}
where the sum runs over all atomic contributions to the angular momentum and the normalization by $N_{\mathrm{cells}}$, the number of unit cells in the model system, gives the angular momentum per unit cell.

In Fig.~\ref{fig:srtio3_circular_drive}(c), we show the angular momentum induced by circularly polarized laser pulses in SrTiO$_{3}$ at \SI{300}{\K}. While the right-handed laser pulse induces an angular momentum parallel to the $z$ axis, the left-handed laser pulse induces the opposite angular momentum. By varying the orientation of the circularly polarized laser pulse, the angular momentum can be aligned along different axes (Fig. S14; SM Note 7~\cite{supplementalMaterial}). Although the total angular momentum of the crystal is conserved, the angular momentum per unit cell is not a conserved quantity~\footnote{Analogous to the difference between momentum and crystal momentum, angular momentum has a crystal counterpart known as pseudoangular momentum~\cite{streibDifferenceAngularMomentum2021}, which is conserved within each unit cell.} and, consequently, after the sharp initial rise to $|L_{z}| \approx \SI{12}{\angstrom\squared\amu\per\ps}$ at \SI{0.2}{\ps}, we observe a steady decrease back to zero for both types of pulses at longer times. This loss of angular momentum is a consequence of the fact that the other modes that the pair of driven modes couples to carry no angular momentum.

\begin{figure}[htbp]
    \includegraphics{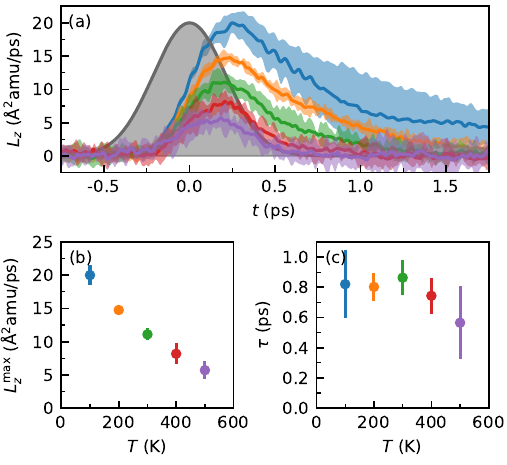}
    \caption{Temperature-dependence of laser-driven dynamics of SrTiO$_{3}$. (a) Time-dependence of generated angular momentum along the $z$ axis in circularly driven SrTiO$_{3}$ and (b) peak angular momentum along the $z$ axis. (c) Lifetimes of the driven phonon modes. Dynamics and lifetimes are obtained from five simulations. Shaded areas and error bars indicated the 95\% ($2 \sigma$) confidence intervals. The envelope of the laser pulse is indicated in gray.}
    \label{fig:srtio3_temperatures}
\end{figure}

Our method also allows us to analyze how temperature affects the generation and decay of angular momentum, by simulating the circular driving across a range of temperatures. The total angular momentum of circularly driven SrTiO$_{3}$ is shown for temperatures ranging from \SI{100}{\K} to \SI{500}{\K} in Fig.~\ref{fig:srtio3_temperatures}. As the temperature increases, the maximum amount of angular momentum generated along the $z$ axis $L_{z}^{\mathrm{max}}$ decreases, and it also decays to zero more rapidly [Fig.~\ref{fig:srtio3_temperatures}(a),(b)]. We attribute the weaker response of the driven modes to enhanced coupling between phonon modes at elevated temperatures. This increased coupling is also reflected in the reduced phonon lifetimes, which decrease above \SI{300}{\K}, going from \SI{0.8}{\ps} down to \SI{0.6}{\ps} at elevated temperatures [Fig.~\ref{fig:srtio3_temperatures}(c)], and in a shift of the peak of the angular momentum to earlier times, as the driven modes reach their steady state sooner as a result of the stronger coupling. 

\section*{Discussion}

In summary, we have demonstrated the utility of laser-driven MD simulations for simulating strongly driven phonons and their nonlinear effects in materials. Our method, implemented in the openly available \texttt{LuMD} code~\cite{lumdCode2026}, can drive both IR- and Raman-active modes in MD simulations, through laser pulses with arbitrary frequency, intensity, orientation, and polarization (linear and circular). This flexibility makes it possible to systematically explore how the mode dynamics depend on the laser frequency and field intensity. Moreover, because MD naturally incorporates temperature effects, the effect of temperature on the mode dynamics is automatically included.

In contrast to the phenomenological models previously used to study nonlinear phononics in materials, the phonon interactions are inherently captured by the used interatomic potentials. As a result of this, our method requires no prior knowledge of the exact interactions between modes, making it directly applicable to any insulating material for which an accurate interatomic potential is available.

Although our analysis focuses on phonon modes at the zone center ($\Gamma$-point), the use of supercells in the simulations offers a natural means to extend the approach to phonon modes at finite wave vectors. By mapping the atomic trajectories onto phonon eigenvectors at arbitrary $\mathbf{q}$-points, one can investigate energy transfer across reciprocal space. This could provide direct insight into the complex, multi-step scattering and coupling processes that govern the dynamics of strongly driven phonons. As such, the method could provide useful insights for phonon conversion processes, such as phonon parametric downconversion~\cite{teitelbaumDirectMeasurementAnharmonic2018}.

\section*{Acknowledgements}

We thank Seongjoo Jung for helping test the code. This work was supported by ETH Z\"{u}rich, the Swiss National Science Foundation (SNF) under Grant No. 225790, and the Dutch Research Council (NWO) under Grant No. 019.253EN.013. Calculations were performed on the Swiss National Supercomputing Center (CSCS) Daint cluster under Project No. lp61 and on the ETH Z\"{u}rich Euler cluster.

\section*{Data availability}

The \texttt{LuMD} code~\cite{lumdCode2026} used to compute mode activities and set up and analyze the laser-driven molecular dynamics simulations, as well as the data that support the findings of this article will become available upon publication.

\bibliography{references}

\end{document}

\title{Supplemental Material: \\ Nonlinear phononics through molecular dynamics simulations}

\author{Mike Pols}
\email{mike.pols@mat.ethz.ch}
\affiliation{%
    Materials Theory, ETH Zurich, CH-8093 Z\"urich, Switzerland
}%

\author{Nicola A. Spaldin}
\email{nicola.spaldin@mat.ethz.ch}
\affiliation{%
    Materials Theory, ETH Zurich, CH-8093 Z\"urich, Switzerland
}%

\date{\today}

\maketitle

\clearpage

\tableofcontents

\clearpage

\section{Laser-induced forces}

\subsection{Infrared (IR) atomic force}

The potential energy associated with a permanent dipole moment $\mathbf{p}$, for example induced by atomic displacements, in the presence of an electric field $\mathbf{E}$ is given by
\begin{equation}
    U^{\mathrm{IR}} = - \mathbf{p} \cdot \mathbf{E}
\end{equation}
and the force on atom $\kappa$ that follows from this is defined as
\begin{equation}
    \mathbf{F}_{\kappa}^{\mathrm{IR}} = - \frac{\partial U^{\mathrm{IR}}}{\partial \mathbf{r}_{\kappa}} = \mathbf{Z}_{\kappa}^{*} \mathbf{E},
\end{equation}
where we use the definition of the Born effective charge tensor, $\mathbf{Z}^{*}_{\kappa} = \frac{\partial \mathbf{p}}{\partial \mathbf{r}_{\kappa}} = \frac{\partial \mathbf{F}_{\kappa}}{\partial \mathbf{E}}$. The Born effective charges can be calculated for a material using density functional perturbation theory (DFPT) and they can be symmetrized using the symmetry operations.

\subsection{Raman atomic force}

For an induced dipole moment $\mathbf{p} = \alpha \mathbf{E}$, the corresponding potential energy, $U^{\mathrm{Raman}}$, in the presence of an electric field $\mathbf{E}$ is given by
\begin{equation}
    U^{\mathrm{Raman}} = - \int^{\mathbf{p}}_{0} \mathbf{E} \cdot d \mathbf{p} = - \int^{\mathbf{E}}_{0} \mathbf{E} \cdot d ( \alpha \mathbf{E} ) = - \frac{1}{2} \mathbf{E}^{T} \alpha \mathbf{E}.
\end{equation}
The corresponding atomic force then follows from the derivative with respect to the atomic position of this expression, as
\begin{equation}
    \mathbf{F}_{\kappa}^{\mathrm{Raman}} = - \frac{\partial U^{\mathrm{Raman}}}{\partial \mathbf{r}_{\kappa}} = \frac{1}{2} \mathbf{E}^{T} \left( \frac{\partial \alpha}{\partial \mathbf{r}_{\kappa}} \right) \mathbf{E},
\end{equation}
where we introduce the atomic Raman tensor $\frac{\partial \alpha}{\partial \mathbf{r}_{\kappa}}$.

\subsection{Implementation in LAMMPS}

LAMMPS~\cite{thompsonLAMMPSFlexibleSimulation2022} is a molecular dynamics (MD) code with a high degree of flexibility. Our implementation for laser driving materials uses the modular structure of the code. In a standard MD simulation, the energies, forces, and stresses are evaluated from the interatomic potential at every timestep. These quantities are then used to propagate the system iteratively in time until the desired time window for the material dynamics of interest is covered.

\begin{figure}[htbp]
    \includegraphics{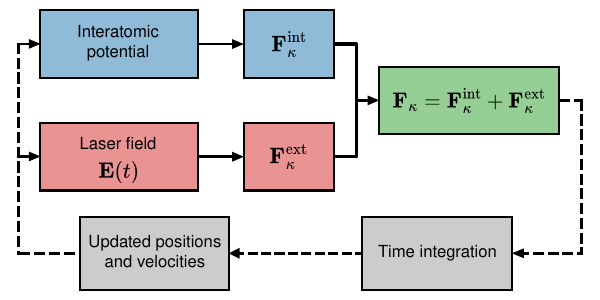}
    \caption{Schematic of molecular dynamics simulation with \texttt{LuMD}. At each time step, the internal forces $\mathbf{F}^{\mathrm{int}}_{\kappa}$ are evaluated from the interatomic potential (blue) and the external laser-driven forces $\mathbf{F}^{\mathrm{ext}}_{\kappa}$ are computed (red) and combined into a total atomic force $\mathbf{F}_{\kappa}$ (green). The total forces are then used to propagate the system through time in \texttt{LAMMPS} (gray).}
    \label{si-fig:lumd_schematic}
\end{figure}

To drive the material with the laser in our simulation, we follow a similar approach. However, instead of only using the internal forces from the interatomic potential, we also add an external force, given by Eq. (3), to all atoms to model the effective interaction with the electric field of the laser pulse. In our implementation, which we release as the \texttt{LuMD} code~\cite{lumdCode2026}, the external atomic forces $\mathbf{F}^{\mathrm{ext}}_{\kappa}$ are added to the internal forces $\mathbf{F}^{\mathrm{int}}_{\kappa}$ once the latter have been computed. In doing so, the total force acting on each atom $\kappa$ becomes $\mathbf{F}_{\kappa} = \mathbf{F}^{\mathrm{int}}_{\kappa} + \mathbf{F}^{\mathrm{ext}}_{\kappa}$, which is then use to propagate the system through time. Through this scheme the interactions of a material with a laser pulse can be treated without modifying the underlying interatomic potential or time-integration method. As a result, the implementation is fully compatible with the existing functionality of LAMMPS. A schematic overview of the simulation scheme with and without a laser pulse is shown in Fig.~\ref{si-fig:lumd_schematic}.

\clearpage

\section{Atomic Raman tensors}

\subsection{First-principles calculations}

The atomic Raman tensors are not straightforwardly available in standard density functional theory (DFT) codes. However, we note that the polarizability of a unit cell is related to the dielectric tensor of the material as $\alpha = \epsilon_{0} V ( \epsilon^{\infty} - \mathbf{I} )$, with $\epsilon_{0}$ the vacuum permittivity, $V$ the unit cell volume, $\epsilon^{\infty}$ the macroscopic dielectric tensor, and $\mathbf{I}$ the identity matrix. As such, we can write the atomic Raman tensor as a derivative of the dielectric tensor, as
\begin{equation}
    \frac{\partial \alpha}{\partial \mathbf{r}_{\kappa}} = \epsilon_{0} V \frac{\partial \epsilon^{\infty}}{\partial \mathbf{r}_{\kappa}}.
\end{equation}

Here, we use a finite-difference method with respect to the atomic positions to calculate the atomic Raman tensors. We displace each of the symmetry unique atoms within the material along the three Cartesian directions, i.e. $x$, $y$, and $z$ axes, and compute the macroscopic dielectric tensor for these displaced structures with density functional perturbation theory (DFPT). Along with the dielectric tensor of the equilibrium structure, we use the dielectric tensors of these displaced structures to fit the atomic Raman tensors, which we symmetrize using the symmetry operations of the material.

\subsection{Convergence with respect to atomic displacements}

\begin{figure}[htbp]
    \includegraphics{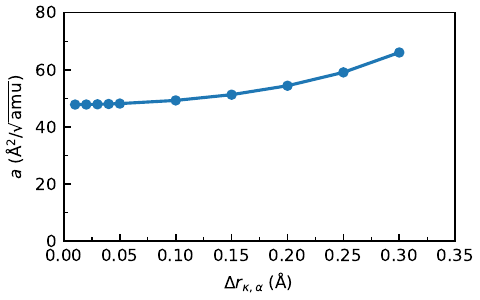}
    \caption{Convergence of Raman tensor with respect to the step size of the Cartesian atomic displacements. We show the dependence of $a$, which is the nonzero component in the Raman tensor of the $xy$-polarized mode at \SI{39.2}{\THz}.}
    \label{si-fig:atomic_raman_tensor_convergence}
\end{figure}

Next, we check the convergence of the atomic Raman tensors with respect to the size of the Cartesian atomic displacements in diamond. We do so by calculating the Raman tensor of the mode at \SI{39.2}{\THz} using Eq.(9) for displacements ranging from \SI{0.01}{\angstrom} to \SI{0.30}{\angstrom}. The $xy$ component of this mode, has a Raman tensor with the following form
\begin{equation}
    \mathbf{R} = 
    \begin{pmatrix}
        0 & a & 0 \\
        a & 0 & 0 \\
        0 & 0 & 0 \\
    \end{pmatrix},
\end{equation}
allowing us to track the coefficient $a$ as a function of the step size to check for convergence. The result of this analysis is shown in Figure~\ref{si-fig:atomic_raman_tensor_convergence}. For large displacements, we observe a quadratic dependence of the mode Raman tensor on the step size, which we attribute to nonlinear higher-order terms rather than the atomic Raman tensor. Since the Raman tensor of the mode, and thus also the atomic Raman tensors, levels off for displacements below \SI{0.05}{\angstrom}, we choose $\Delta r_{\kappa, \alpha} = \pm\SI{0.03}{\angstrom}$ ($\alpha = x, y, z$)
to compute the atomic Raman tensors for Raman-based excitation in diamond.

\clearpage

\section{Machine-learning interatomic potentials}

\subsection{Generating datasets}

To create a dataset that describes the out-of-equilibrium driving of materials, structurally diverse structures have to be added to the dataset during the on-the-fly molecular dynamics (MD) simulations. We characterize the local atomic environments during these simulations using an adaptation of the smooth overlap of atomic positions (SOAP) descriptor~\cite{bartokRepresentingChemicalEnvironments2013}. The two-body radial and three-body angular descriptors are defined with cutoff radii of \SI{8.0}{\angstrom} and \SI{5.0}{\angstrom}, respectively. Atomic positions are smoothed via Gaussian distributions with a width of \SI{0.5}{\angstrom}. Both descriptors are expanded in a basis set of spherical Bessel functions and Legendre polynomials, employing 12 and 8 Bessel functions for the two-body and three-body terms, respectively, with a maximum angular momentum quantum number of $l_{\mathrm{max}} = 3$. The resulting expansion coefficients form the descriptor for the local atomic environments. To quantify the structural similarity and distinguish between diverse atomic configurations, we use a polynomial kernel of power 4, where the two-body radial and three-body angular descriptors are weighted by 0.1 and 0.9, respectively.

\begin{table*}[htbp!]
\caption{\label{tab:mlip_datasets}Datasets for different materials.}
\begin{ruledtabular}
    \begin{tabular}{lcccccc}
    Material & $T_{1}$ (\SI{}{\K}) & $N_{\mathrm{DFT},1}$ (-) & $T_{1}$ (\SI{}{\K}) & $N_{\mathrm{DFT},2}$ (-) & $T_{1}$ (\SI{}{\K}) &  $N_{\mathrm{DFT},3}$ (-) \\ \colrule
    LiNbO$_{3}$     & 500   & 168   & 1000  & 311   & 1500  & 511   \\
    SrTiO$_{3}$     & 500   & 164   & 1000  & 229   & 1500  & 305   \\
    Diamond (C)     & 500   & 29    & 1500  & 72    & 2500  & 109   \\
    \end{tabular}
\end{ruledtabular}
\end{table*}

For all materials, supercells are used to generate the datasets for training the machine-learning interatomic potentials (MLIPs). To avoid artifacts arising from non-orthogonal periodic boundary conditions, supercells with orthogonal lattice vectors are employed for all systems, resulting in supercells containing 360 atoms for LiNbO$_{3}$, 320 atoms for SrTiO$_{3}$, and 512 atoms for diamond. The datasets for LiNbO$_{3}$ and SrTiO$_{3}$ are generated at lower temperatures than the diamond dataset, reflecting the lower-frequency vibrational modes present in these materials. Each on-the-fly MD simulation is performed for \SI{100.0}{\ps} using a time step of \SI{1.0}{\fs}. During these simulations, the temperature and pressure are controlled through Parrinello-Rahman dynamics~\cite{parrinelloCrystalStructurePair1980, parrinelloPolymorphicTransitionsSingle1981} with friction coefficients $\gamma$ = \SI{5.0}{\per\ps} and $\gamma_{\mathrm{L}}$ = \SI{5.0}{\per\ps} for the atomic and lattice degrees of freedom, respectively. An overview of the simulation temperatures and resulting dataset sizes is provided in Table~\ref{tab:mlip_datasets}.

\subsection{Model accuracy}

\subsubsection{Training and validation errors}

As we split the dataset into a training set and validation set, we can evaluate the accuracy of the MLIPs against both these subsets of data. The errors of the models for the training and validation sets are shown in Table~\ref{si-tab:mlip_errors}. We see that the MLIPs are highly accurate in terms of energies, forces, and stresses for all materials.

\begin{table*}[htbp!]
	\caption{\label{si-tab:mlip_errors}Errors of the machine-learning interatomic potentials (MLIPs) for the training and validation sets.}
	\begin{ruledtabular}
		\begin{tabular}{lcccc}
			Model & Dataset & $\Delta_{E}$ (\SI{}{\meV}/atom) & $\Delta_{F}$ (\SI{}{\meV\per\angstrom}) & $\Delta_{\sigma}$ (\SI{}{\meV\per\angstrom\cubed}) \\ \colrule
			LiNbO$_{3}$     & Training		& 0.1   & 19.5  & 0.2   \\
			                & Validation	& 0.1   & 23.5  & 0.1   \\ \hline
			Diamond (C)     & Training		& 0.2   & 16.9  & 2.3   \\
			                & Validation	& 0.2   & 20.3  & 2.3   \\ \hline
			SrTiO$_{3}$     & Training		& 0.6   &  9.0  & 0.8   \\
			                & Validation	& 0.6   & 10.2  & 0.9   \\
		\end{tabular}
	\end{ruledtabular}
\end{table*}

\subsubsection{Phonon dispersion}

\begin{figure}[htbp]
    \includegraphics{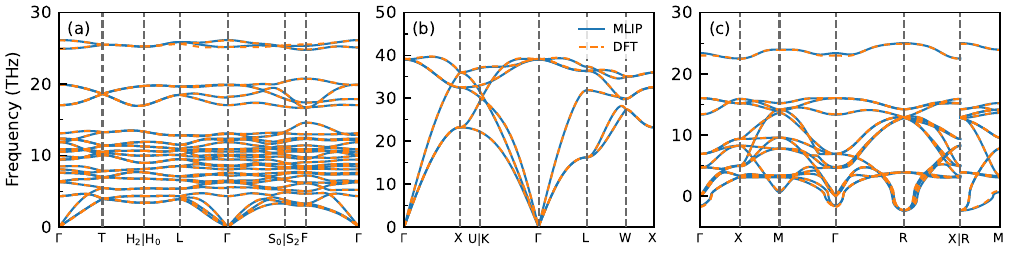}
    \caption{Comparison of phonon dispersions calculated with DFT and the trained MLIPs. Phonon dispersions are shown for (a) LiNbO$_{3}$, (b) diamond, and (c) SrTiO$_{3}$. The MLIP-calculated dispersion is shown in blue, while the DFT-calculated dispersion is shown in dashed orange lines.}
    \label{si-fig:mlip_phonons}
\end{figure}

It is important for the dynamical simulations of the phonons that the MLIPs reproduce the phonons with first-principles accuracy. To validate that our models have this accuracy, we compare the phonon dispersons as computed with DFT with those calculated using the MLIPs. For both methods, we compute the phonon dispersions using the finite-displacement method as implemented in \texttt{phonopy}~\cite{togoFirstprinciplesPhononCalculations2023, togoImplementationStrategiesPhonopy2023}. We apply atomic displacements of \SI{0.01}{\angstrom} in the supercells of size $3 \times 3 \times 3$ (LiNbO$_{3}$), $5 \times 5 \times 5$ (diamond), and $4 \times 4 \times 4$ (SrTiO$_{3}$). We account for long-range dipole-dipole interactions between phonons through the non-analytical correction term~\cite{gonzeInteratomicForceConstants1994, gonzeDynamicalMatricesBorn1997}, using the Born effective charges and dielectric tensor calculated with DFPT for both the DFT- and MLIP-based phonon dispersion. The resulting phonon dispersions are shown in Fig.~\ref{si-fig:mlip_phonons}. We highlight that the MLIP reproduces the DFT-based phonon dispersion well, with minimal shifts in the phonon frequencies, indicating the models are suited to study phonon dynamics.

\subsubsection{Phonon anharmonicity}

Ultimately, if we want to study nonlinear processes in materials, we should also validate how well the MLIPs capture the anharmonicity of the vibrational modes of a material. While generally accepted that MLIPs, when given the appropriate training data, will capture the anharmonicity of the material~\cite{verdiThermalTransportPhase2021, ouyangAccurateDescriptionHighorder2022}, here we here validate how well the MLIP for LiNbO$_{3}$ reproduces the anharmonicity of the material. An indirect validation of how well the MLIP for SrTiO$_{3}$ captures its anharmonicity is shown in Note~\ref{si-note:strontium_titanate}, where we validate how the frequency of the ferroelectric soft mode changes as a function of temperature in Fig.~\ref{si-fig:srtio3_natural_frequency}.

For LiNbO$_{3}$, we test the MLIP by explicitly calculating the mode anharmonicities and mode couplings using the frozen-phonon method. Specifically, we displace the equilibrium structure along the phonon eigenvectors, with mode amplitudes ranging from from $Q = - \SI{1.0}{\AA \sqrt{\amu}}$ to $Q = \SI{1.0}{\AA \sqrt{\amu}}$ on a grid of 11 equally space points per mode, and evaluate the energies of the resulting structures. For a single mode this leads to 11 structures to be evaluated, whereas the displacement of two and three modes results in 121 and 1331 structures, respectively. We then fit the potential energy surface (PES) as a function of the mode amplitudes to extract the corresponding anharmonic coefficients and mode couplings, such as $a_{\nu \mu \lambda} Q_{\nu} Q_{\mu} Q_{\lambda}$. The phonon eigenvectors obtained from the two methods may differ by an overall factor $-1$. The corresponding coefficients transform accordingly, such that their signs remain consistent with the eigenvector convention. Therefore, we compare the absolute values of the coefficients of the two methods. The resulting coefficients obtained from DFT and the MLIP are shown in Table~\ref{si-tab:mlip_anharmonicity}. We observe that the MLIP very accurately describes the anharmonicity and coupling of the modes, leading to almost identical values between the two methods.

\begin{table*}[htbp!]
	\caption{\label{si-tab:mlip_anharmonicity}Mode anharmonicities and coupling coefficients for $A_{1}$ modes in LiNbO$_{3}$ computed using DFT and the MLIP.}
	\begin{ruledtabular}
		\begin{tabular}{ccccc}
			Mode $\nu$ & Mode $\mu$ & Mode $\lambda$ & $| a^{\mathrm{DFT}}_{\nu \mu \lambda} |$ (\SI{}{\meV\per\AA\cubed\per\sqrt{\amu}\cubed}) & $| a^{\mathrm{MLIP}}_{\nu \mu \lambda} |$ (\SI{}{\meV\per\AA\cubed\per\sqrt{\amu}\cubed}) \\ \colrule
             9 &  9 &  9 &   18.98 &   18.64 \\
			12 & 12 & 12 &   18.26 &   17.66 \\
            16 & 16 & 16 &    6.84 &    6.65 \\
            27 & 27 & 27 &  169.20 &  166.94 \\ \hline
             9 &  9 & 12 &   18.58 &   17.40 \\
             9 & 12 & 12 &   45.98 &   45.43 \\ \hline
             9 &  9 & 16 &    2.49 &    1.90 \\
             9 & 16 & 16 &   17.90 &   17.83 \\ \hline
             9 &  9 & 27 &   26.19 &   25.53 \\
             9 & 27 & 27 &  138.83 &  137.73 \\ \hline
            12 & 12 & 16 &   24.30 &   24.87 \\
            12 & 16 & 16 &    2.83 &    2.69 \\ \hline
            12 & 12 & 27 &    0.03 &    0.84 \\
            12 & 27 & 27 &   77.86 &   77.03 \\ \hline
            16 & 16 & 27 &   94.55 &   93.41 \\
            16 & 27 & 27 &   79.59 &   82.24 \\ \hline
             9 & 12 & 16 &   29.40 &   29.43 \\
             9 & 12 & 27 &   56.93 &   54.83 \\
             9 & 16 & 27 &    0.92 &    4.15 \\
            12 & 16 & 27 &   76.51 &   78.72 \\
		\end{tabular}
	\end{ruledtabular}
\end{table*}

\clearpage

\section{Laser pulses}

\subsection{Laser fluence}

The fluence of a laser pulse can be obtained by integrating the instantaneous intensity over time. For a pulse propagating in vacuum, the intensity is related to the electric field through $I ( t ) = \epsilon_{0} c \, | E ( t ) |^{2}$, giving
\begin{equation}
    \label{si-eq:laser_fluence}
    F_{\mathrm{laser}} = \int^{+\infty}_{-\infty} I ( t ) \, dt = \int^{+\infty}_{-\infty} \epsilon_{0} c \, | E ( t ) |^{2} \, dt = \epsilon_{0} c \frac{E_{0}^{2}}{2} \sqrt{\pi} \sigma \left[ 1 + e^{-\omega_{0}^{2} \sigma^{2}} \right],
\end{equation}
where $\sigma = \tau / \sqrt{8 \ln 2}$ is the Gaussian pulse width expressed in terms of the full width at half maximum (FWHM) of the pulse envelope. In the limit of many optical cycles within the pulse envelope, $\omega_{0} \sigma \gg 1$, the oscillatory contribution averages out and the fluence reduces to $F_{\mathrm{laser}} \approx \frac{E_{0}^{2}}{2} \sqrt{\pi} \sigma$. Conversely, in the limit of a pulse shorter than a single optical cycle, $\omega_{0} \sigma \ll 1$, the carrier envelope oscillation is negligible, resulting in $F_{\mathrm{laser}} \approx E_{0}^{2} \sqrt{\pi} \sigma$.

\subsection{Laser properties}

Here, we provide an overview of the pulse parameters and pulse properties of the laser pulses that we use to drive phonon dynamics in the example materials. The central frequency, peak electric field $E_{0}$, pulse width $\tau$,  and laser fluence $F_{\mathrm{laser}}$ can all be found in Table~\ref{tab:laser_parameters}.

\begin{table*}[htbp!]
\caption{\label{tab:laser_parameters}Laser parameters used for laser driving the various materials.}
\begin{ruledtabular}
    \begin{tabular}{cccccc}
    Driving & Material & Frequency (\SI{}{\THz}) & $E_{0}$ (\SI{}{\mega\volt\per\cm}) & $\tau$ (\SI{}{\ps}) & $F_{\mathrm{laser}}$ (\SI{}{\milli\joule\per\cm\squared}) \\ \colrule
    IR              & LiNbO$_{3}$   &  18.2     & 8.0       & 0.40  &  25.57    \\
    IR              & SrTiO$_{3}$   &   3.0     & 2.0       & 0.50  &   2.00    \\ \colrule
    Raman (SFE)     & Diamond       &  19.5     & 32.0      & 0.20  & 204.59    \\
    Raman (ISRS)    & Diamond       & 759.0     & 128.0     & 0.01  & 163.67    \\ 
    \end{tabular}
\end{ruledtabular}
\end{table*}

\subsection{Spectral range}

\begin{figure}[htbp]
    \includegraphics{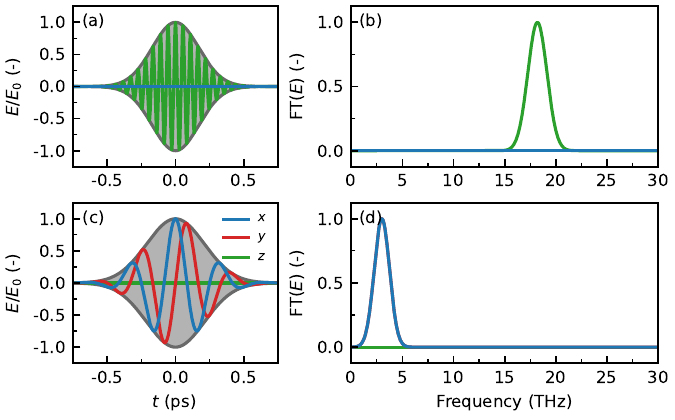}
    \caption{Laser pulses used for IR driving. (a) Electric field components and (b) Fourier transform of the electric field components of the \SI{18.2}{\THz} laser pulse used to drive LiNbO$_{3}$. (c) Electric field components and (d) Fourier transform of the electric field components of the \SI{3.0}{\THz} laser pulse used to circularly drive SrTiO$_{3}$. The $x$, $y$, and $z$ components are indicated in blue, red, and green, respectively.}
    \label{si-fig:ir_laser_pulses}
\end{figure}

\begin{figure}[htbp]
    \includegraphics{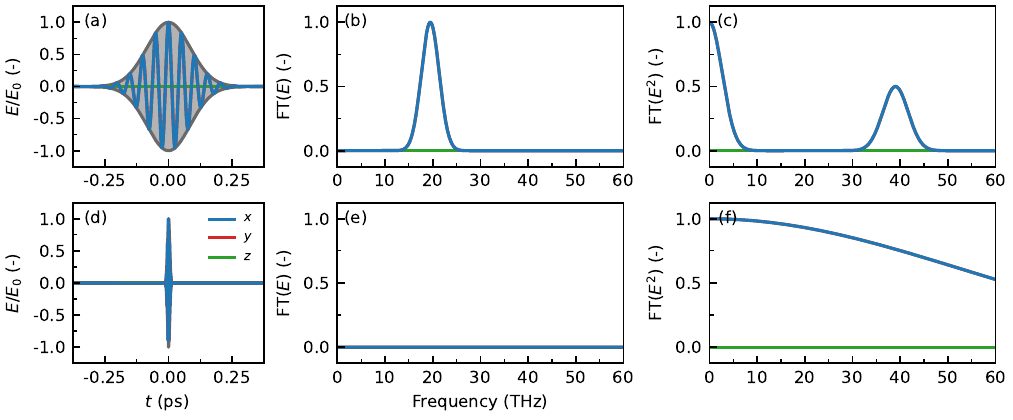}
    \caption{Laser pulses used for Raman driving. (a) Electric field components, (b) Fourier transform of the electric field components, and (c) Fourier transform of the square of the electric field components used to drive diamond with SFE. (d) Electric field components, (e) Fourier transform of the electric field components, and (f) Fourier transform of the square of the electric field components used to drive diamond with ISRS. The $x$, $y$, and $z$ components are indicated in blue, red, and green, respectively.}
    \label{si-fig:raman_laser_pulses}
\end{figure}

For insights into the spectral range of the laser pulses used in our simulations, we show in Fig.~\ref{si-fig:ir_laser_pulses} and Fig.~\ref{si-fig:raman_laser_pulses} the time profiles of the electric fields together with their spectral components. Looking at the spectral range of the laser pulses used for IR driving [Fig.~\ref{si-fig:ir_laser_pulses}], we see that the spectral range of the laser pulse used for driving LiNbO$_{3}$ overlaps with the IR-active mode of interest at \SI{18.2}{\THz} [Fig.~\ref{si-fig:ir_laser_pulses}(b)]. Similarly, we see that the laser pulse used to circularly drive SrTiO$_{3}$ also overlaps with the IR-active mode found at \SI{3.0}{\THz}. When we instead look at the laser pulses used for Raman driving [Fig.~\ref{si-fig:raman_laser_pulses}], we see that the pulse used to drive the Raman-active mode of diamond at \SI{39}{\THz} via SFE is centered at \SI{19.5}{\THz} [Fig.~\ref{si-fig:raman_laser_pulses}(b)]. Since the driving force for SFE is proportional to the square of the electric field, the second harmonic of this pulse is found at \SI{39}{\THz}, which overlaps with the Raman-active mode of interest [Fig.~\ref{si-fig:raman_laser_pulses}(c)]. The pulse used for ISRS driving is very short [Fig.~\ref{si-fig:raman_laser_pulses}(d)], which results in a very broad spectral range that overlaps with the \SI{39}{\THz} mode, allowing it to be driven impulsively.

\clearpage

\section{Lithium niobate}

\subsection{System sizes}

\begin{figure}[htbp]
    \includegraphics{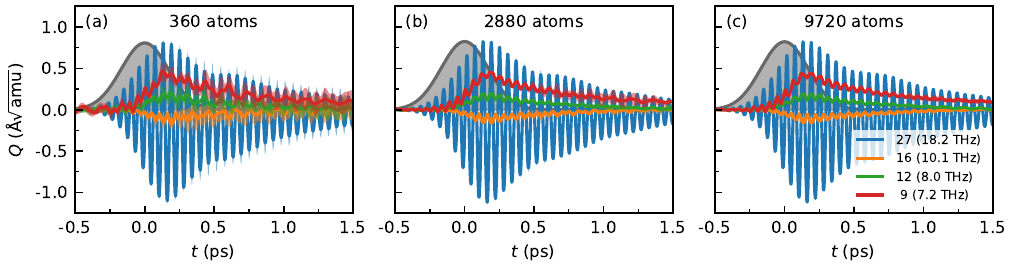}
    \caption{Convergence of mode dynamics of LiNbO$_{3}$ with respect to system size. Mode dynamics of $A_{1}$ modes in LiNbO$_{3}$ for a system of (a) 360 atoms, (b) 2880 atoms, and (c) 9720 atoms. The dynamics from MD are the average dynamics from 5 simulations, with the 95\% ($2 \sigma$) confidence interval for the mode amplitudes indicated in shaded ares. The envelope of the laser pulse is indicated in gray.}
    \label{si-fig:linbo3_system_sizes}
\end{figure}

Next, we check the convergence of the laser-driven MD simulations for LiNbO$_{3}$ with respect to the size of the model systems. To do so, we drive the highest-frequency $A_{1}$ mode in LiNbO$_{3}$ for increasing system sizes, using a $z$-oriented laser pulse with central frequency \SI{18.2}{\THz} ($E_{0} = \SI{8.0}{\mega\volt\per\cm}$; $\tau = \SI{0.4}{\ps}$) at \SI{50}{\K}. For every system size, i.e. 360, 2880, and 9720 atoms, we run a total of ten independent simulations. Figure~\ref{si-fig:linbo3_system_sizes} shows the ensemble-averaged dynamics of the $A_{1}$ modes for the three different system sizes. As the size of the model systems increases, the mode dynamics progressively converge. Given the small uncertainty in the mode amplitudes for the 2880-atom model system, we consider the dynamics to be sufficiently converged at this system size.

\subsection{Damping constants}

\begin{figure}[htbp]
    \includegraphics{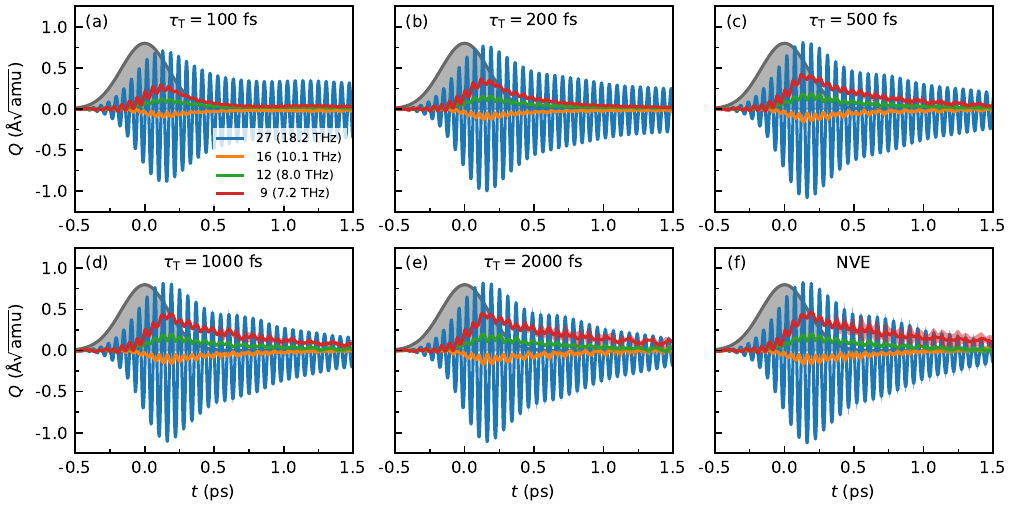}
    \caption{Effect of thermostat on mode dynamics of LiNbO$_{3}$. Mode dynamics of $A_{1}$ modes in LiNbO$_{3}$ with a damping constant for the temperature ($\tau_{\mathrm{T}}$) of (a) \SI{100}{\fs}, (b) \SI{200}{\fs}, (c) \SI{500}{\fs}, (d) \SI{1000}{\fs}, and (e) \SI{2000}{\fs}. (f) The laser-driven dynamics in an isolated system, i.e. $NVE$ simulation. The dynamics from MD are the average dynamics from 5 simulations, with the 95\% ($2 \sigma$) confidence interval for the mode amplitudes indicated in shaded ares. The envelope of the laser pulse is indicated in gray.}
    \label{si-fig:linbo3_damping_constants}
\end{figure}

To assess whether the thermostat that we apply during the laser driving affects the overall mode dynamics of LiNbO$_{3}$, we compare the dynamics with and without an applied thermostat during laser excitation. As before, we focus on the dynamics resulting from driving the high-frequency $A_{1}$ mode. Specifically, we drive the mode in an $NVT$ ensemble with damping constants for the thermostat $\tau_{\mathrm{T}}$ of \SI{100}{\fs}, \SI{200}{\fs}, \SI{500}{\fs}, \SI{1000}{\fs}, and \SI{2000}{\fs}, as well as in an $NVE$ ensemble at \SI{50}{\K}. The resulting ensemble-averaged mode dynamics are shown in Fig.~\ref{si-fig:linbo3_damping_constants}. We observe that systems coupled to a weak thermostat, i.e. those with large values of $\tau_{\mathrm{T}}$, exhibit dynamics similar to those of the $NVE$ ensemble. In contrast, for more strongly coupled thermostats with $\tau_{\mathrm{T}} \leq \SI{200}{\fs}$, the mode dynamics are noticeably affected. In particular, the fluctuations observed for modes 9, 12, and 16 in the $NVE$ ensemble, are suppressed in simulations with strongly coupled thermostats, indicating that the temperature damping is too aggressive.

\begin{figure}[htbp]
    \includegraphics{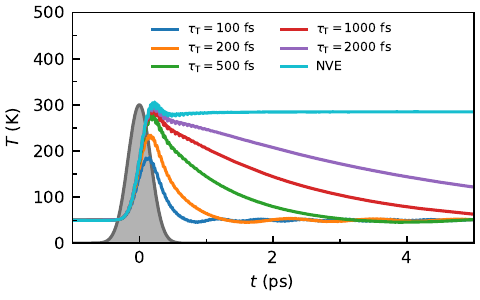}
    \caption{Time evolution of temperature in laser-driven LiNbO$_{3}$ for an $NVE$ ensemble and for $NVT$ ensembles with different damping constants $\tau_{\mathrm{T}}$ for the temperature. Temperatures are averaged over 5 simulations and window-averaged over \SI{200}{\fs}. The envelope of the laser pulse is indicated in gray.}
    \label{si-fig:linbo3_temperatures}
\end{figure}

For additional insights into the effects of a thermostat on laser-driven systems, we also look into the time evolution of the temperature for these simulations. The time evolution of the temperature in these simulations is shown in Fig.~\ref{si-fig:linbo3_temperatures}. From the $NVE$ simulation we see that the laser excitation heats up the model system by about \SI{250}{\K} to approximately \SI{300}{\K}. Consistent with the mode dynamics, the temperature evolution also shows that strongly coupled thermostats can lead to unphysical dynamics. In these cases, the temperature increase is substantially reduced, with the temperature even decreasing below the initial temperature, following the laser excitation. For larger damping constants, the initial temperature increase closely follows that of the $NVE$ simulation, indicating a minimal impact of the thermostat on the laser heating. At longer times, however, such weakly-coupled thermostats gradually decrease the temperature of the system, mimicking the dissipation of energy to the environment that real materials would experience. This suggests that a large damping constant, such as $\tau_{\mathrm{T}} = \SI{2000}{\fs}$, provides a sufficiently weak coupling that preserves the initial laser-induced dynamics while still allowing for heat dissipation at longer times.

\subsection{Phonon lifetimes}

We make use of the time autocorrelation functions of the mode amplitudes to extract the lifetimes of the phonon modes. The autocorrelation function is defined as
\begin{equation}
    \label{si-eq:autocorrelation_function}
    C ( \tau ) = \langle Q ( t ) Q ( t + \tau ) \rangle,
\end{equation}
where $Q ( t )$ is the mode amplitude at time $t$, $\tau$ is the time lag, and the brackets $\langle \ldots \rangle$ denote a time average. To be able to determine how quickly the autocorrelation function decays independently of its magnitude, we normalize it as
\begin{equation}
    \label{si-eq:normalized_autocorrelation_fuction}
    \phi ( \tau ) = \frac{\langle Q ( t ) Q ( t + \tau ) \rangle}{\langle Q ( t )^{2} \rangle} = \frac{C ( \tau )}{C ( 0 )}
\end{equation}
which by definition has $\phi ( 0 ) = 1$. 

\begin{figure}[htbp]
    \includegraphics{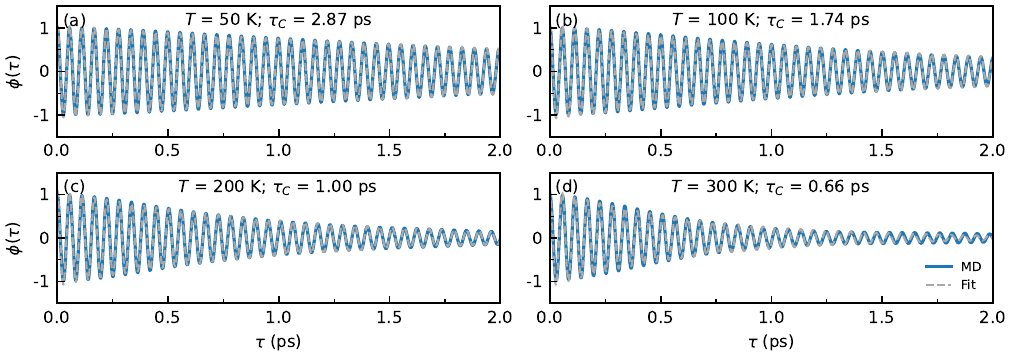}
    \caption{Autocorrelation function of the \SI{18.2}{\THz} mode in LiNbO$_{3}$ following laser excitation with a laser pulse with peak electric field of $E_{0} = \SI{2.0}{\mega\volt\per\cm}$. Investigated temperatures shown are (a) \SI{50}{\K}, (b) \SI{100}{\K}, (c) \SI{200}{\K}, and (d) \SI{300}{\K}. The autocorrelation functions each represent the data from a single simulation, the actual phonon lifetimes are computed by averaging over various simulations in the ensemble.}
    \label{si-fig:linbo3_autocorrelation}
\end{figure}

The lifetime of the mode is subsequently obtained by fitting a damped cosine function of the form
\begin{equation}
    \label{si-eq:damped_consine}
    \phi ( \tau ) = A \exp \left( - \frac{\tau}{\tau_{C}} \right) \cos ( \omega \tau )
\end{equation}
where $A$ is a scaling constant, $\tau_{C}$ is the phonon lifetime also called the correlation time, and $\omega$ is the angular frequency of the mode. In practice, the fit is performed over the first \SI{2.0}{\ps} of the autocorrelation function. This procedure is performed independently for each MD run, and the resulting phonon lifetimes are obtained by averaging over the ensemble of runs with the same conditions. Figure~\ref{si-fig:linbo3_autocorrelation} shows representative autocorrelation functions and their corresponding damped cosine fits for the \SI{18.2}{\THz} mode in LiNbO$_{3}$, driven by a laser pulse with a peak electric field of $E_{0} = \SI{2.0}{\mega\volt\per\cm}$. We observe a stronger decay for higher temperatures, indicating a shorter lifetime of the driven mode at elevated temperatures.

\clearpage

\section{Diamond}

\subsection{System sizes}

\begin{figure}[htbp]
    \includegraphics{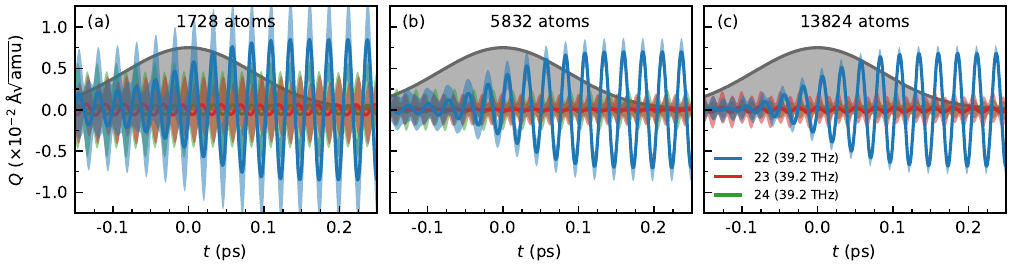}
    \caption{Convergence of mode dynamics of diamond with respect to system size. Mode dynamics of $F_{\mathrm{2g}}$ modes in diamond for a system of (a) 1728 atoms, (b) 5832 atoms, and (c) 13824 atoms. The dynamics from MD are the average dynamics from 10 simulations, with the 95\% ($2 \sigma$) confidence interval for the mode amplitudes indicated in shaded ares. The envelope of the laser pulse is indicated in gray.}
    \label{si-fig:diamond_system_sizes}
\end{figure}

Analogous to LiNbO$_{3}$, we also check the convergence of the laser-driven dynamics of diamond with respect to the size of the model system. In this case, we drive the triply-degenerate $F_{\mathrm{2g}}$ Raman-active modes at approximately \SI{39.2}{\THz} via sum-frequency excitation (SFE), using an $xy$-oriented laser pulse with a central frequency of \SI{19.5}{\THz} ($E_{0} = \SI{8.0}{\mega\volt\per\cm}$). A total of ten simulations are performed for system sizes of 1728, 5832, and 13824 atoms. The ensemble-average dynamics resulting from this excitation are shown in Fig.~\ref{si-fig:diamond_system_sizes}. As is the case for LiNbO$_{3}$, we see that the laser-driven dynamics converges with larger system sizes. We deem the dynamics of the $F_{\mathrm{2g}}$ modes to be sufficiently converged for model systems with 5832 atoms [Fig.~\ref{si-fig:diamond_system_sizes}(b)].

\subsection{Pulse orientation}

\begin{figure}[htbp]
    \includegraphics{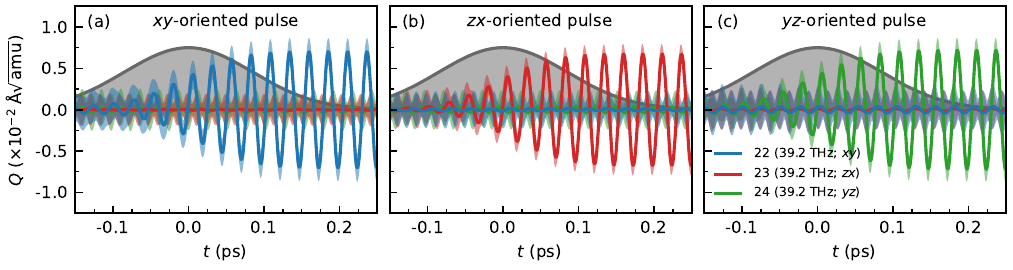}
    \caption{Influence of pulse orientation on mode dynamics of diamond. Mode dynamics of $F_{\mathrm{2g}}$ modes in diamond for an (a) $xy$- (b) $xz$-, and (c) $yz$-oriented laser pulse. The dynamics from MD are the average dynamics from 10 simulations, with the 95\% ($2 \sigma$) confidence interval for the mode amplitudes indicated in shaded ares. The envelope of the laser pulse is indicated in gray.}
    \label{si-fig:diamond_pulse_orientation}
\end{figure}

As a result of its crystal symmetry (space group $Fd\bar{3}m$; point group $O_{h}$), diamond has a triply degenerate Raman-active mode with $F_{\mathrm{2g}}$ symmetry. The three degenerate components can be represented by the basis functions $xy$, $xz$, and $yz$, corresponding to the off-diagonal components of the Raman tensor. In our simulations, these components correspond to mode 22 ($xy$), mode 23 ($xz$), and mode 24 ($yz$). By rotating the orientation of the laser pulse, used for the driving, in space, we can selectively excite a different degenerate Raman-active mode. As before, we use laser pulses with a central frequency of \SI{19.5}{\THz} to drive the material via SFE. The ensemble-averaged orientation-dependent mode dynamics obtained from ten independent simulations are shown in Figure~\ref{si-fig:diamond_pulse_orientation}. As the orientation of the laser pulse is rotated, different components of the degenerate Raman modes are selectively excited.

\clearpage

\section{Strontium titanate}\label{si-note:strontium_titanate}

\subsection{System sizes}

\begin{figure}[htbp]
    \includegraphics{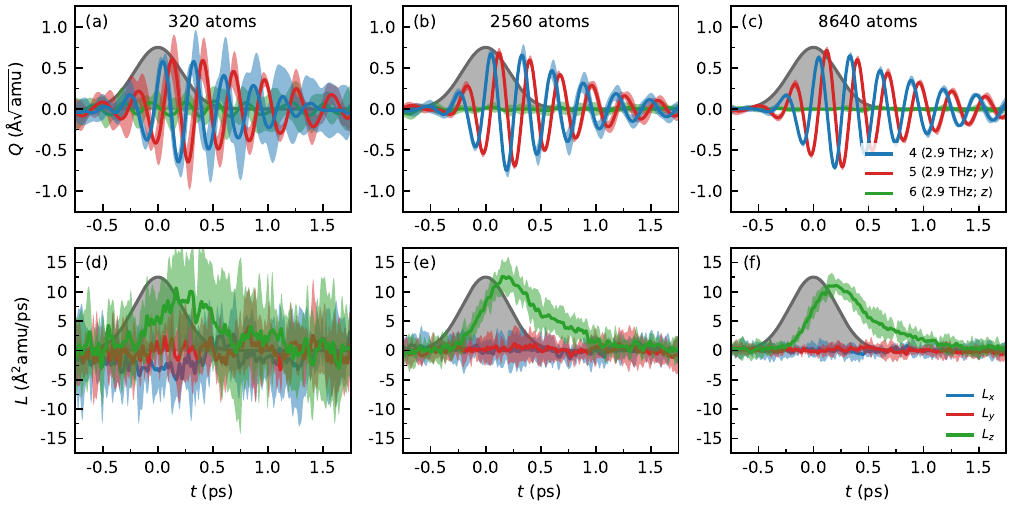}
    \caption{Convergence of laser-driven dynamics of SrTiO$_{3}$ with respect to system size. (a-c) Mode dynamics of $T_{1u}$ modes and (d-f) angular momentum in SrTiO$_{3}$ at \SI{300}{\K} for systems of 320 atoms, 2560 atoms, and 8640 atoms. The dynamics from MD are the average dynamics from 5 simulations, with the 95\% ($2 \sigma$) confidence interval for the mode amplitudes indicated in shaded ares. The envelope of the laser pulse is indicated in gray.}
    \label{si-fig:srtio3_system_sizes}
\end{figure}

Finally, we also check the convergence of the mode dynamics of SrTiO$_{3}$ with respect to the model system size. To do so, we drive the triply degenerate zone-center $T_{1u}$ soft mode at \SI{300}{\K} using a laser pulse with a central frequency of \SI{3.0}{\THz} ($E_{0} = \SI{2.0}{\mega\volt\per\cm}$; $\tau = \SI{0.5}{\ps}$), matching the frequency of this mode at this temperature. We run five independent simulations for systems with 320, 2560, and 8640 atoms. In Fig.~\ref{si-fig:srtio3_system_sizes}, we show the resulting ensemble-averaged mode dynamics of the $T_{1u}$ modes [Fig.~\ref{si-fig:srtio3_system_sizes}(a-c)] and angular momentum [Fig.~\ref{si-fig:srtio3_system_sizes}(d-f)]. Analogous to laser-driven LiNbO$_{3}$ and diamond, we see that the mode dynamics converges as the model system size is increased. Given the noisier nature of the angular momentum of the system, we deem the laser-driven material dynamics to be converged for model systems of 8640 atoms.

\subsection{Mode frequency}

\begin{figure}[htbp]
    \includegraphics{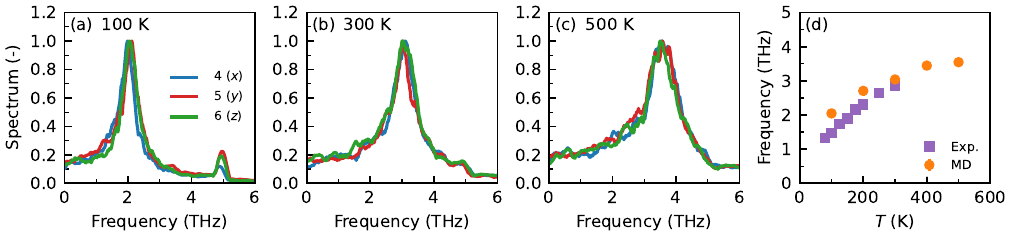}
    \caption{Temperature dependence of $T_{1u}$ mode frequency in SrTiO$_{3}$. Spectral components of the modes at (a) \SI{100}{\K}, (b) \SI{300}{\K}, and (c) \SI{500}{\K}. (d) Average frequency of the three $T_{1u}$ modes. A Savitzky-Golay filter with a window size of 25 and polynomial order of 3 is applied to smooth the spectral components.}
    \label{si-fig:srtio3_natural_frequency}
\end{figure}

To investigate the temperature dependence of the of the $T_{1u}$ mode frequencies, we analyze the spectral components of the mode amplitudes at finite temperatures. We perform 5 independent MD simulations at temperatures ranging from \SI{100}{\K} to \SI{500}{\K}, with each individual simulation spanning \SI{50}{\ps}. The spectral characteristics of the $T_{1u}$ modes are obtained by Fourier transforming the mode amplitudes from each simulation. In Fig.~\ref{si-fig:srtio3_natural_frequency}(a)-(c), we show the spectral components averaged over the five independent simulations. The mode frequencies are determined from the peak position of the spectral components. We find that the frequency of the modes increases as the temperature is increased; while at \SI{100}{\K} the modes have a frequency of about \SI{2.0}{\THz}, this increases to approximately \SI{3.0}{\THz} and \SI{3.7}{\THz} at \SI{300}{\K} and \SI{500}{\K}, respectively. As shown in Fig.~\ref{si-fig:srtio3_natural_frequency}(d), the observed mode frequencies match well with experimental data for the temperature dependence of the soft mode in SrTiO$_{3}$~\cite{sirenkoSoftmodeHardeningSrTiO32000}.

\subsection{Pulse orientation}

\begin{figure}[htbp]
    \includegraphics{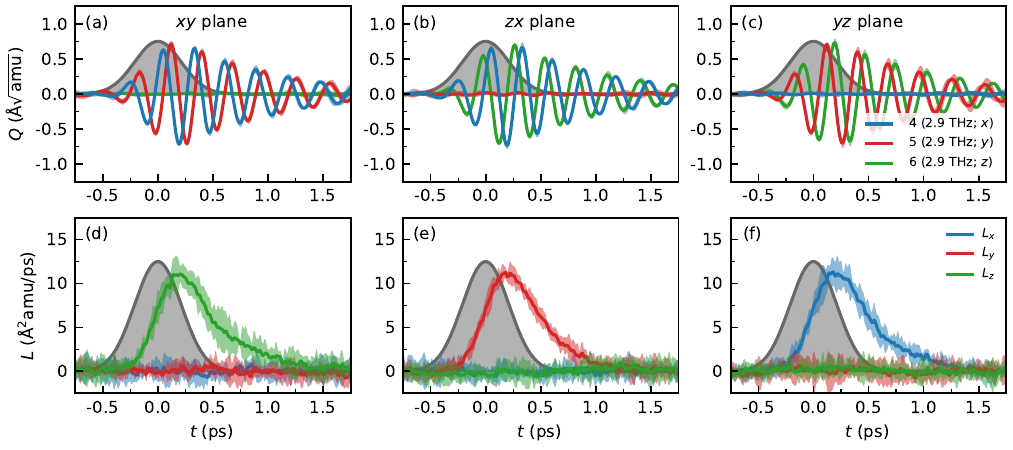}
    \caption{Effect of pulse orientation on generated angular momentum in SrTiO$_{3}$. (a-c) Mode dynamics of $T_{1u}$ modes and (d-f) generated angular momentum components in SrTiO$_{3}$ for laser pulses that are circularly polarized within the $xy$, $zx$, and $yz$ plane, respectively. The dynamics from MD are the average dynamics from 5 simulations, with the 95\% ($2 \sigma$) confidence interval for the mode amplitudes indicated in shaded area. The envelope of the laser pulse is indicated in gray.}
    \label{si-fig:srtio3_pulse_orientation}
\end{figure}

In the main text, SrTiO$_{3}$ is driven by a laser pulse that is circularly polarized within the $xy$ plane. The angular momentum that results from this pulse is aligned with the $z$ axis ($L_{z} \neq 0$), while the $x$ and $y$ components vanish, $L_{x} = L_{y} = 0$. By changing the orientation of the laser pulse, different components of the $T_{1u}$ modes can be excited and the orientation of the resulting angular momentum can be controlled. In Fig.~\ref{si-fig:srtio3_pulse_orientation}, we show the dynamics obtained for laser pulses that are circularly polarized within the $xy$, $zx$, and $yz$ planes. These results confirm that different laser pulse orientations result in different orientations for the generated angular momentum. Specifically, the angular momentum is always perpendicular to the plane in which the laser pulse is circularly polarized.

\clearpage

\bibliography{references}